\documentclass[pra,aps,showpacs,superscriptaddress,preprint]{revtex4}
\usepackage{amsmath,bm}
\usepackage{amsfonts}
\usepackage{amssymb}
\usepackage{graphicx}
\usepackage{dcolumn}
\usepackage{bm}

\newcommand{\therm}{thermalization}
\begin{document}
\title{Quantum quench dynamics of the Bose-Hubbard model at finite temperatures}

\author{J.~M. Zhang}
\affiliation{Beijing National Laboratory for Condensed
Matter Physics, Institute of Physics, Chinese Academy of
Sciences, Beijing 100080, China} \affiliation{FOCUS Center
and MCTP, Department of Physics, University of Michigan,
Ann Arbor, Michigan 48109, USA}
\author{C. Shen}
\affiliation{FOCUS Center and MCTP, Department of Physics,
University of Michigan, Ann Arbor, Michigan 48109, USA}
\author{W.~M. Liu}
\affiliation{Beijing National Laboratory for Condensed
Matter Physics, Institute of Physics, Chinese Academy of
Sciences, Beijing 100080, China}

\begin{abstract}
We study quench dynamics of the Bose-Hubbard model by exact
diagonalization. Initially the system is at thermal
equilibrium and of a finite temperature. The system is then
quenched by changing the on-site interaction strength $U$
suddenly. Both the single-quench and double-quench
scenarios are considered. In the former case, the
time-averaged density matrix and the real-time evolution
are investigated. It is found that though the system
thermalizes only in a very narrow range of the quenched
value of $U$, it does equilibrate or relax well in a much
larger range. Most importantly, it is proven that this is
guaranteed for some typical observables in the
thermodynamic limit. In order to test whether it is
possible to distinguish the unitarily evolving density
matrix from the time-averaged (thus time-independent),
fully decoherenced density matrix, a second quench is
considered. It turns out that the answer is affirmative or
negative according to the intermediate value of $U$ is zero
or not.
\end{abstract}
 \maketitle

\section{Introduction}\label{sec1}

Out-of-equilibrium dynamics following a quantum quench is a
topic of intense study at present. The theme is pursued
primarily along two lines. The first one is about the
equilibration and \therm\ mechanism of a quantum system
\cite{bloch11,rigol_09,rigol_integrable,rigol_nature,
manmana,roux09,roux2,kollath07,weiss,hanggi}, a fundamental
yet still open issue in statistical physics. The second one
is about the the real-time dynamical behavior of a
many-body system
\cite{greiner,sengupta,luttinger,dimer,polkovnikov}, which
is highly non-trivial in the regime where the
quasi-particle picture breaks down.

Among all the models investigated so far, the Bose-Hubbard
model takes a special position. As a paradigmatic
strongly-correlated model, it can be realized accurately
with cold atoms in optical lattices, and especially, the
parameters can be controlled (e.g. changed suddenly) to a
high degree \cite{zoller,bloch,bloch_np}. This nice
property makes it an ideal candidate to investigate quantum
quench dynamics both theoretically and experimentally. Up
to now, in the few theoretical works on the quench dynamics
of the Bose-Hubbard model
\cite{roux09,roux2,kollath07,dimer,sengupta}, the state of
the system before the quench is always assumed to be the
ground state of the initial Hamiltonian. That is, the
system is assumed to be at zero temperature initially.
However, in this paper we shall start from a thermal
equilibrium state. One should note that this scenario is
actually more experimentally relevant. Because in current
experiments, one generally gets not a single tube of cold
atoms, but instead a two-dimensional array of
one-dimensional lattices for the cold atoms
\cite{bloch_np}. In other words, an \textit{ensemble} of
one-dimensional Bose-Hubbard models is obtained in one
shot. Moreover, in view of the fact that the cold atoms are
at finite temperatures necessarily \cite{qi,zhou}, it is
reasonable to start from a thermal state described by a
canonical ensemble density matrix [see Eq.~(\ref{rhoi})
below].

As emphasized by Linden \textit{et al}.~\cite{linden}, in
the pursuit of \therm, it is important to distinguish the
two closely related but inequivalent concepts of
equilibration and \therm. The latter is much stronger and
has the trademark feature of the Boltzmann distribution,
whereas the former refers only to the stationary property
of the density matrix of a (sub)system or some physical
observables. It is highly possible that a system
equilibrates but without \therm. This is actually the case
for the Bose-Hubbard model. As revealed both in previous
works (zero temperature case) \cite{roux09,kollath07} and
in the present paper (finite temperature case), the
Bose-Hubbard model thermalizes only if the quench amplitude
is not so large, at least at the finite sizes currently
accessible. However, it will be shown below that in a much
wider range of parameters, some generic physical
observables equilibrate very well. Among them are the
populations on the Bloch states, which are ready to measure
by the typical time-of-flight experiment \cite{rmp}.
Remarkably, this is actually guaranteed for these
quantities in the thermodynamic limit, i.e., when the size
of the system gets large enough.

The equilibration behavior of the physical observables
imposes a question. It is ready to recognize that the
equilibration of the physical observables is largely an
effect of interference cancelation. It never means that the
density matrix has suffered any dephasing or decoherence.
Actually, the density matrix evolves unitarily and in the
diagonal representation of the Hamiltonian, its elements
simply rotate at constant angular velocities. A natural
question is then, does the time-dependence of the density
matrix has any chance to exhibit it, given that it is
almost absent in the average values of the physical
observables? This leads us to consider giving the system a
second quench. The concern is, would the system yield
different long-time behaviors if the second quench comes at
different times? It turns out that the answer depends on
whether the intermediate Hamiltonian is integrable or
non-integrable. In the former case, the density matrix
shows repeated appreciable recurrences and thus the
dependence on the second quench time is apparent. In the
latter case, on the contrary, the density matrix shows no
sign of recurrence and quantitatively similar long-time
dynamics is observed for quenches at different times.

This paper is organized as follows. In Sec.~\ref{sec2}, the
setting of the problem and the basic approaches are given.
In Sec.~\ref{sec3}, the dynamics after a single quench is
studied. The time-averaged density matrix and the real-time
evolution of some physical observables are investigated in
detail. Based on the observation in this Section, we
proceed to study the scenario of a second quench in
Sec.~\ref{sec4}. Finally, we summarize the results in
Sec.~\ref{sec5}.

\section{Basic formalism}\label{sec2}
The time-dependent Hamiltonian of the Bose-Hubbard model is
($\hbar=k_B=1$ throughout this paper)
\begin{equation}\label{h}
H(t)=-J\sum_{l=1}^M (a_l^\dagger a_{l+1}+a_{l+1}^\dagger
a_l)+ \frac{U(t)}{2}\sum_{l=1}^M a_l^\dagger a_l^\dagger
a_l a_l.
\end{equation}
Here $M$ is the number of sites (the total atom number will
be denoted as $N$) and $a_l^\dagger$ ($a_l$) is the
creation (annihilation) operator for an atom at site $l$.
Note that here periodic boundary condition is assumed. The
parameters $J$ and $U$ are the nearest-neighbor hopping
strength and the on-site atom-atom interaction strength,
respectively. Note that the dynamics of the system depends
only on the ratio $U/J$, thus we will set $J=1$ throughout.
We say the system is quenched if $U$ is changed suddenly at
some time from one value to another value. Experimentally,
for cold atoms in an optical lattice, this can be realized
by using the Feshbach resonance.

Assume that initially the parameter $U$ is of value $U_i$
(the corresponding Hamiltonian is denoted as $H_i$), and
the system is at thermal equilibrium and of inverse
temperature $\beta_i=1/T_i$. Denote the $m$-th eigenvalue
and eigenstate of $H_i$ as $E_m^{i}$ and
$|\psi_m^{i}\rangle$, respectively. The initial density
matrix of the system is then
\begin{eqnarray} \label{rhoi}
\rho_i=\frac{1}{Z_i}\exp(-\beta_i H_i)=\sum_{m=1}^D p^i_m
|\psi_m^{i}\rangle \langle \psi_m^i |,
\end{eqnarray}
where $Z_i=\sum_{m=1}^D \exp(-\beta_i E_m^i)$ is the
partition function and $p^i_m=\frac{1}{Z_i}\exp(-\beta_i
E^i_m)$ is the probability of occupying the eigenstate
$|\psi_m^{i}\rangle $. Note that
$D=\frac{(M+N-1)!}{(M-1)!N!}$ is the dimension of the
Hilbert space $\mathcal{H}$. The density matrix at time $t$
is given formally as $\rho(t)=U(t)\rho_i U^\dagger(t)$,
with $U(t)=\mathcal{T}\exp[-i\int_0^t d \tau H(\tau)]$.
Here $\mathcal{T}$ means time ordering.

The Hamiltonian $H(t)$ is invariant under the translation
$(a_l,a_l^\dagger)\rightarrow (a_{l+1},a_{l+1}^\dagger)$.
This indicates that the total quasi-momentum of the system
$ q=\sum_{k=0}^{M-1} k a_k^\dagger a_k \pmod M $, where
$a_k^\dagger= \frac{1}{\sqrt{M}}\sum_{l=1}^M \exp(i 2\pi k
l/M) a_l^\dagger$ is the creation operator for an atom in
the $k$-th Bloch state, is conserved. This property implies
that if the full Hilbert space is decomposed into $M$
subspaces according to the values of $q$, i.e.,
$\mathcal{H}=\oplus_{q=0}^{M-1}\mathcal{H}^{(q)} $, the
Hamiltonian and the density matrix are always
block-diagonal with respect to the $q$-subspaces,
i.e.,$H(t)=\oplus_{q=0}^{M-1} H^{(q)}(t)$ and
$\rho(t)=\oplus_{q=0}^{M-1} \rho^{(q)}(t)$ \cite{zjm,zjm2}.
It is then possible to study the dynamics in each subspace
individually (which saves a lot of computational resource)
and then gather the information together (note that for the
expectation values of quantities like $ a_k^\dagger a_k  $,
there are contributions from each subspace). Here it is
necessary to mention that though we should have done the
gathering or averaging process for many quantities studied
below, we would rather not do so, because it is observed
that the system behaves quantitatively similar in all the
$q$-subspaces \cite{beta}. A single $q$-subspace captures
the overall behavior very well. Therefore, our strategy is
to focus on some specific $q$-subspace ($q=1$ actually) and
take the normalization $tr(\rho^{(q)}(t))=1$. \textit{It is
understood that in the following all Hamiltonians, density
matrices, eigenvalues, and eigenstates refer to those
belonging to this specific $q$-subspace. We will drop the
superscript $q$ for notational simplicity.}

\section{A single quench}\label{sec3}
Suppose at time $t=0$ the system is quenched by changing
the value of $U$ from $U_i$ to $U_{f_1}$, which is then
held on forever. The Hamiltonian later will be denoted as
$H_{f_1}$, and the eigenvalues and eigenstates associated
will be denoted as $E^{f_1}_n$ and $|\psi^{f_1}_n \rangle$,
respectivley. In the representation of
$\{|\psi^{f_1}_n\rangle \}$, the density matrix at time $t$
is then simply (in this paper $\langle \cdots \rangle$
means quantum state averaging while $\overline{\cdots}$
means time averaging)
\begin{equation}\label{rhot}
\rho(t)=\sum_{m,n=1}^{D_q} e^{-i(E^{f_1}_m-E^{f_1}_n)t}
\langle \psi^{f_1}_m | \rho_i | \psi^{f_1}_n \rangle |
\psi^{f_1}_m \rangle \langle \psi^{f_1}_n |,
\end{equation}
where $D_q \simeq D/M$ is the dimension of the specific
$q$-subspace. It will prove useful to define the
time-averaged density matrix
\begin{eqnarray} \label{def_bar_rho}
\bar{\rho} &=& \lim_{T \rightarrow \infty} \frac{1}{T}\int_0^{T} dt \rho(t)  \nonumber \\
 &=& \sum_{\substack{m,n=1\\ E^{f_1}_m=E^{f_1}_n}}^{D_q} \langle \psi^{f_1}_m | \rho_i | \psi^{f_1}_n \rangle | \psi^{f_1}_m \rangle \langle \psi^{f_1}_n |.
\end{eqnarray}
The time-averaged density matrix is of great relevance for
our purposes. First, it is both time-independent and
variable-independent. Second, the time-averaged value of an
arbitrary operator $O$ is given simply by $\overline
{\langle O\rangle} \equiv \lim_{T \rightarrow \infty}
\frac{1}{T} \int_0 ^{T}  tr(\rho(t)O)d t=tr(\bar{\rho}O)$.
That is, the time-averaged density matrix contains the
overall information of the dynamics of the system.
Actually, as we will see later, for some quantities which
fluctuate little in time, the time-averaged density matrix
tells almost a complete story. Third, the process of
averaging over time is a process of relaxation in the sense
that the entropy associated with $\bar{\rho}$ is definitely
no less than that with the density matrix at an arbitrary
time, i.e., $S(\bar{\rho})\geq S(\rho(t))=S(\rho_i)$. This
is a corollary of the Klein inequality \cite{chuang} and is
reasonable since $\rho_i$ contains all the information of
$\bar{\rho}$ while the inverse is invalid. The equality
also means that $\rho(t)$ will never be damped, and
time-averaging is essential.

Note that when $U_{f_1}\neq 0$, generally there is no
degeneracy between the eigenvalues of $H_{f_1}$. Therefore
the time-averaged density matrix is simply diagonal in the
basis of $\{|\psi^{f_1}_n \rangle \}$, i.e.,
\begin{eqnarray} \label{diag}
\bar{\rho} &=& \sum_{m=1}^{D_q} \langle \psi^{f_1}_m | \rho_i |\psi^{f_1}_m \rangle | \psi^{f_1}_m \rangle \langle \psi^{f_1}_m | \nonumber \\
&\equiv & \sum_{m=1}^{D_q} p_m | \psi^{f_1}_m \rangle
\langle \psi^{f_1}_m |,
\end{eqnarray}
with
\begin{equation} \label{pm}
p_m= \langle \psi^{f_1}_m | \rho_i | \psi^{f_1}_m \rangle=
\frac{1}{Z_i}\sum_{n=1}^{D_q} e^{-\beta_i E_n^i} |\langle
\psi_n^{i}| \psi_m^{f_1}  \rangle |^2
\end{equation}
being the population on the eigenstate
$|\psi_m^{f_1}\rangle$. In the special case of $U_{f_1}=0$,
the Hamiltonian reduces to
$H_{f_1}=\sum_{k=0}^{M-1}\omega_{k} a^\dagger_k a_k$, with
$\omega_k=-2J\cos (2\pi k/M)$. In this case, each
eigenvalue is of the form $\sum_k n_k \omega_k$, under the
constraints $\sum_k n_k=N$ and $\sum_k k n_k \equiv q \pmod
M$, and there can be level degeneracy. However, we can
always make some unitary transforms in each degenerate
subspace to make sure that $\bar{\rho}$ is in the form of
(\ref{diag}).

\subsection{Time-averaged density matrix}

\begin{figure*}[tb]
\begin{minipage}[b]{0.30 \textwidth}
\centering
\includegraphics[ width=\textwidth]{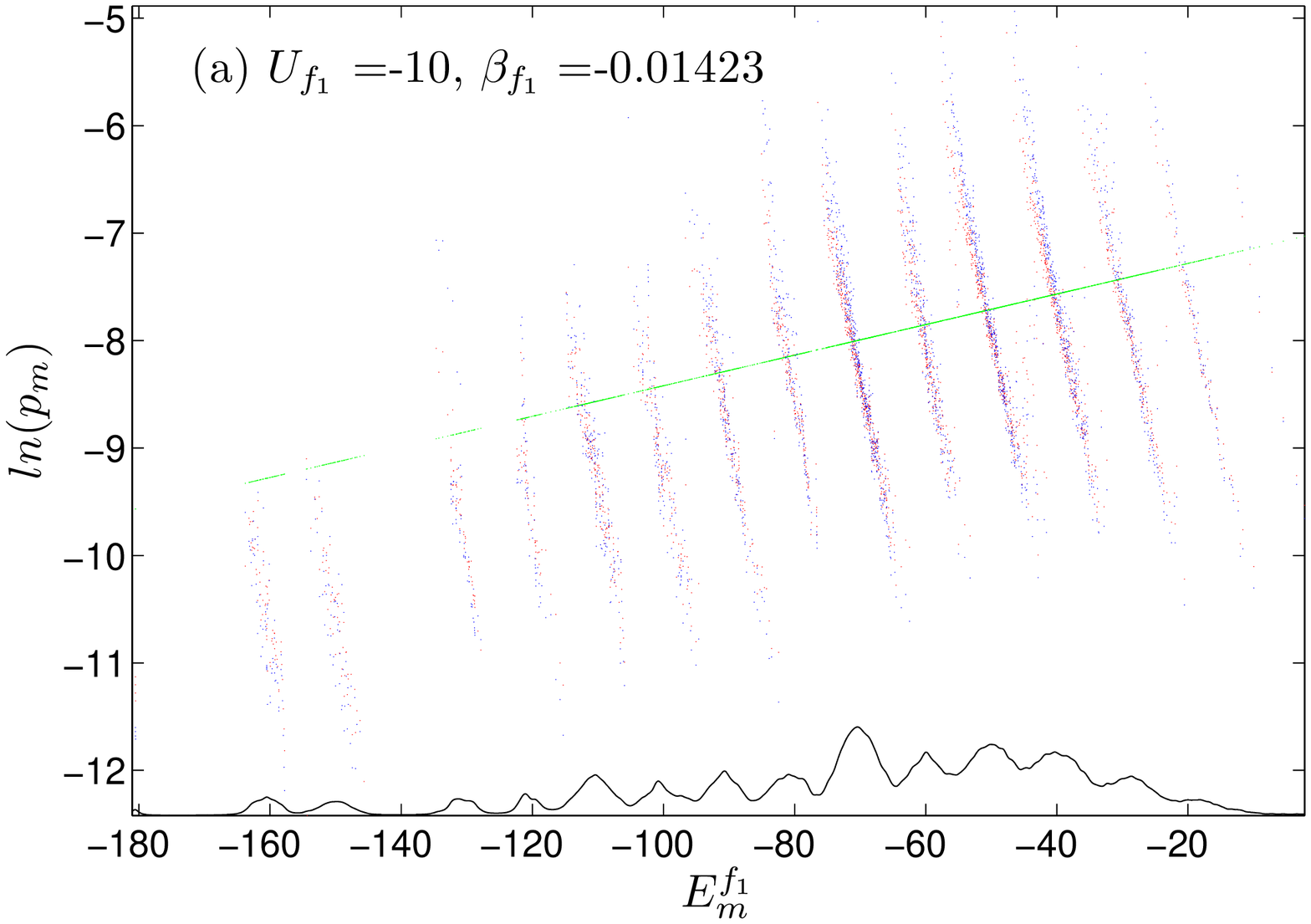}
\end{minipage}
\begin{minipage}[b]{0.30 \textwidth}
\centering
\includegraphics[ width=\textwidth]{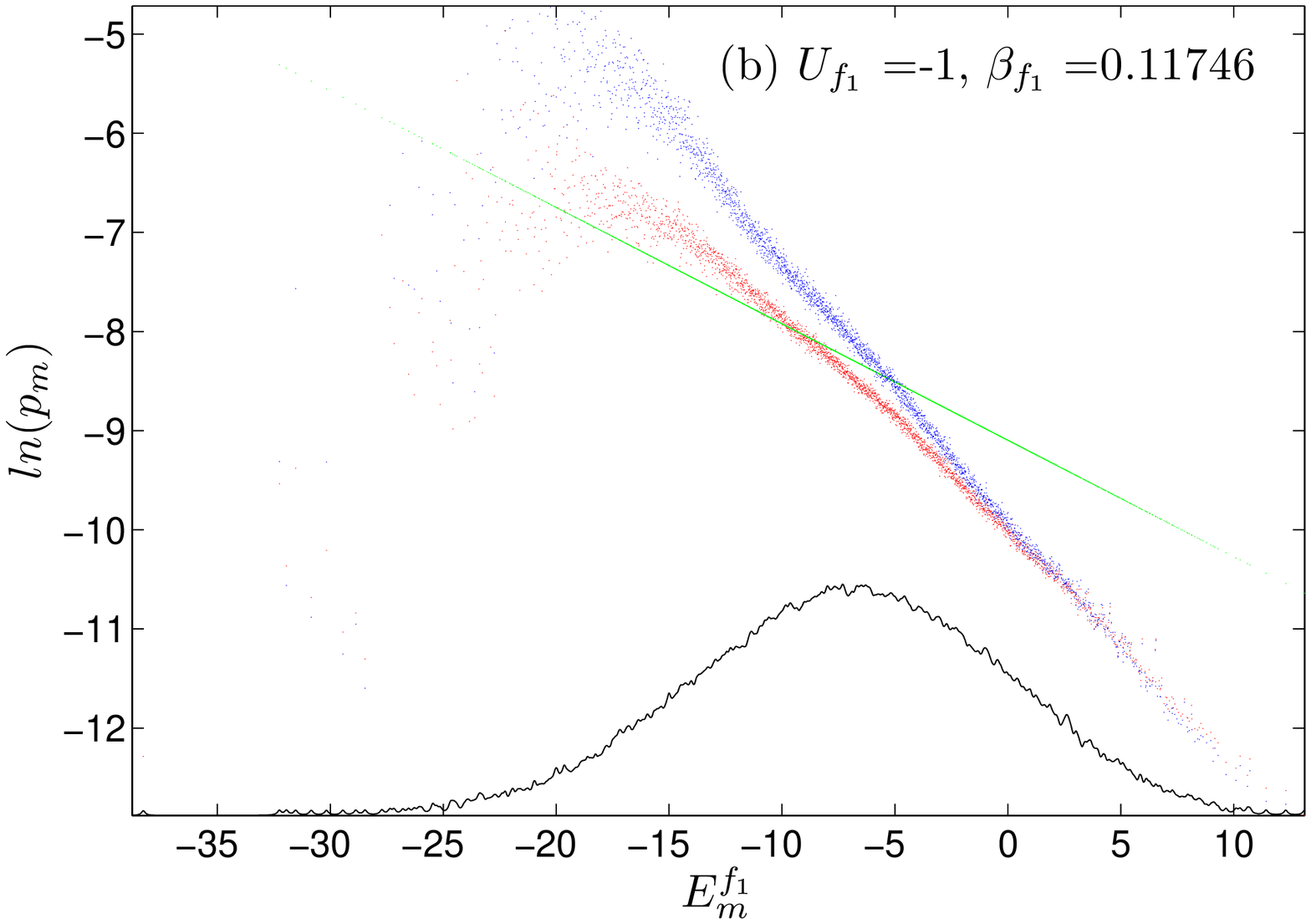}
\end{minipage}
\begin{minipage}[b]{0.30 \textwidth}
\centering
\includegraphics[ width=\textwidth]{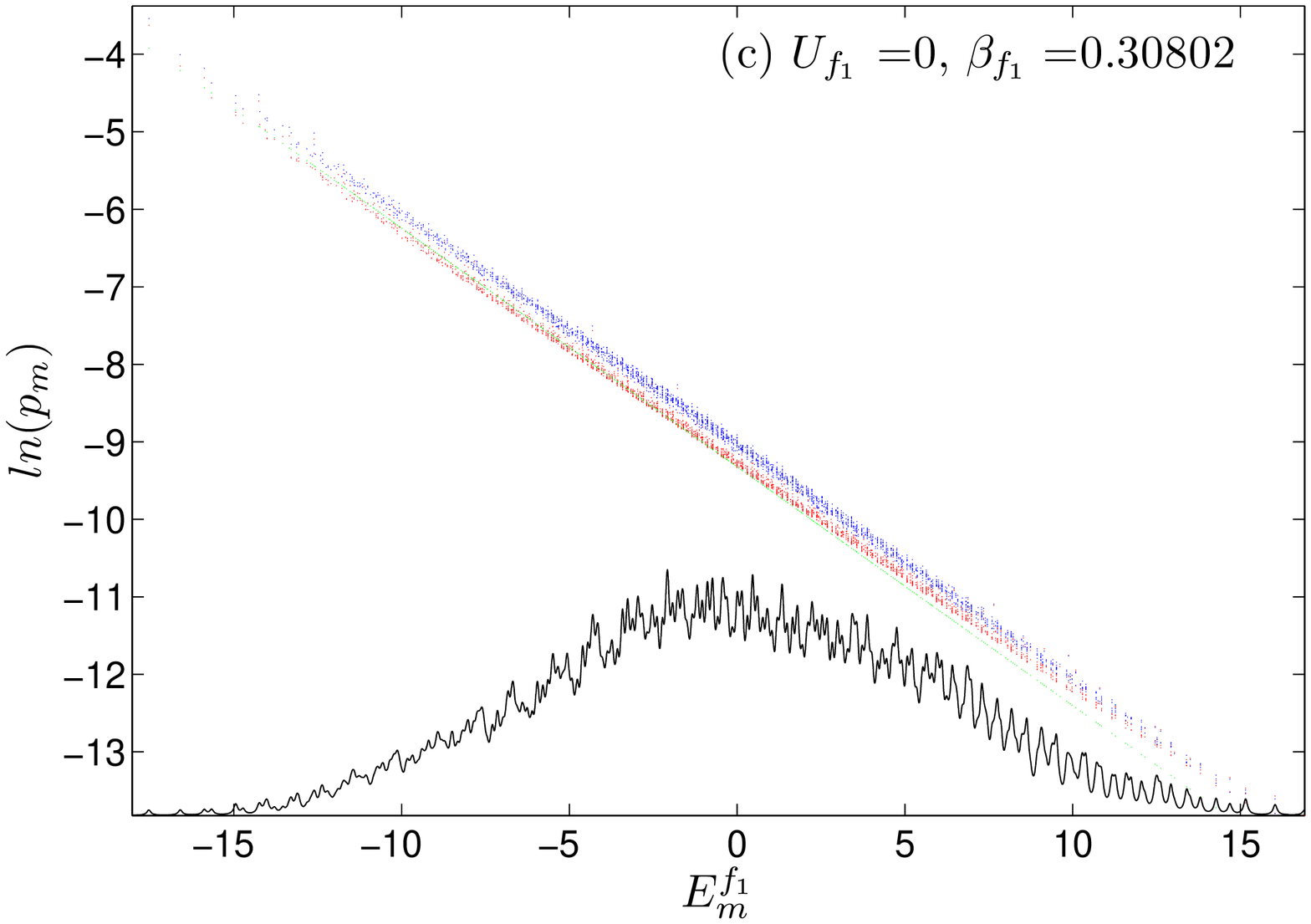}
\end{minipage}

\begin{minipage}[b]{0.30 \textwidth}
\centering
\includegraphics[ width=\textwidth]{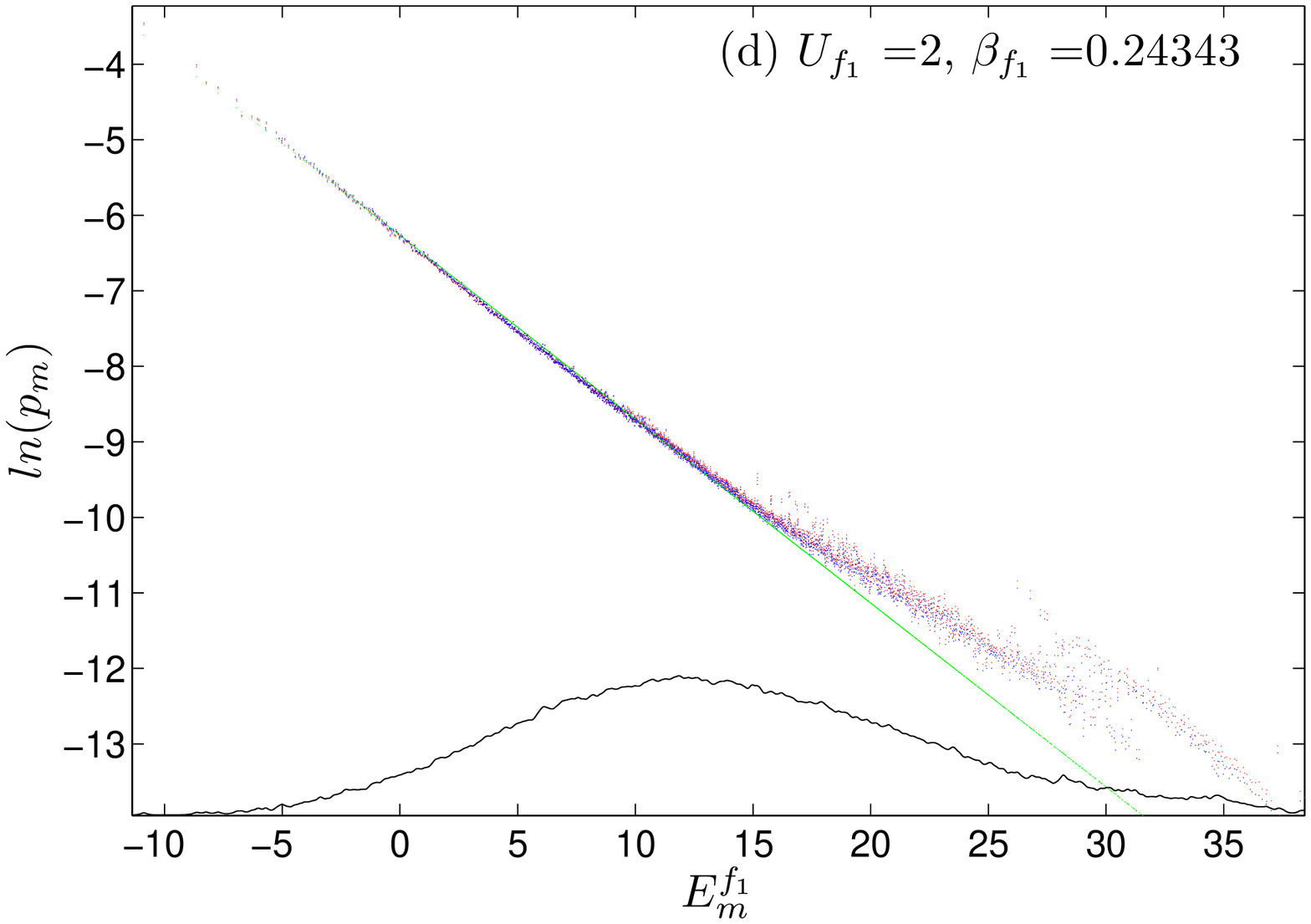}
\end{minipage}
\begin{minipage}[b]{0.30 \textwidth}
\centering
\includegraphics[ width=\textwidth]{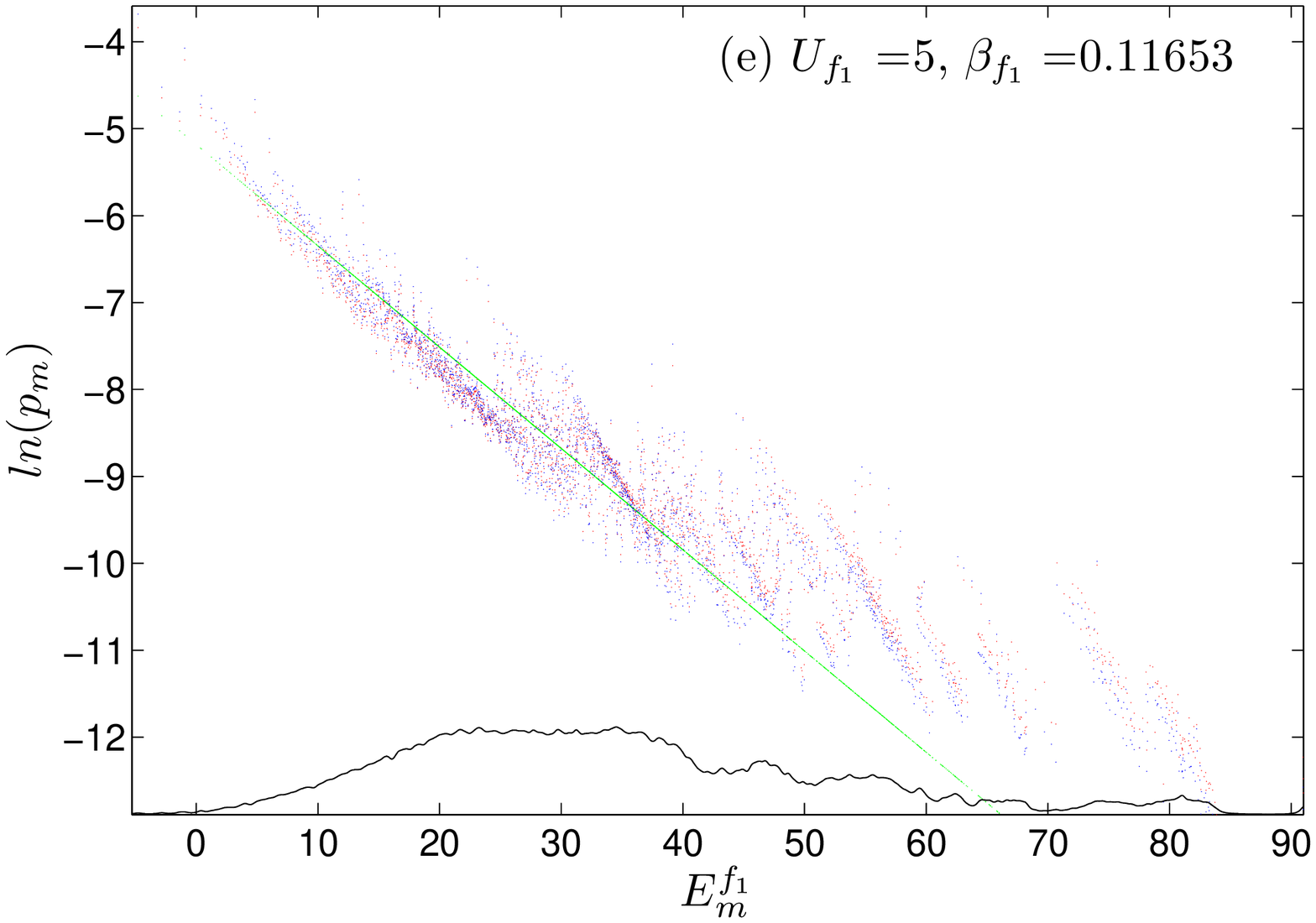}
\end{minipage}
\begin{minipage}[b]{0.30 \textwidth}
\centering
\includegraphics[ width=\textwidth]{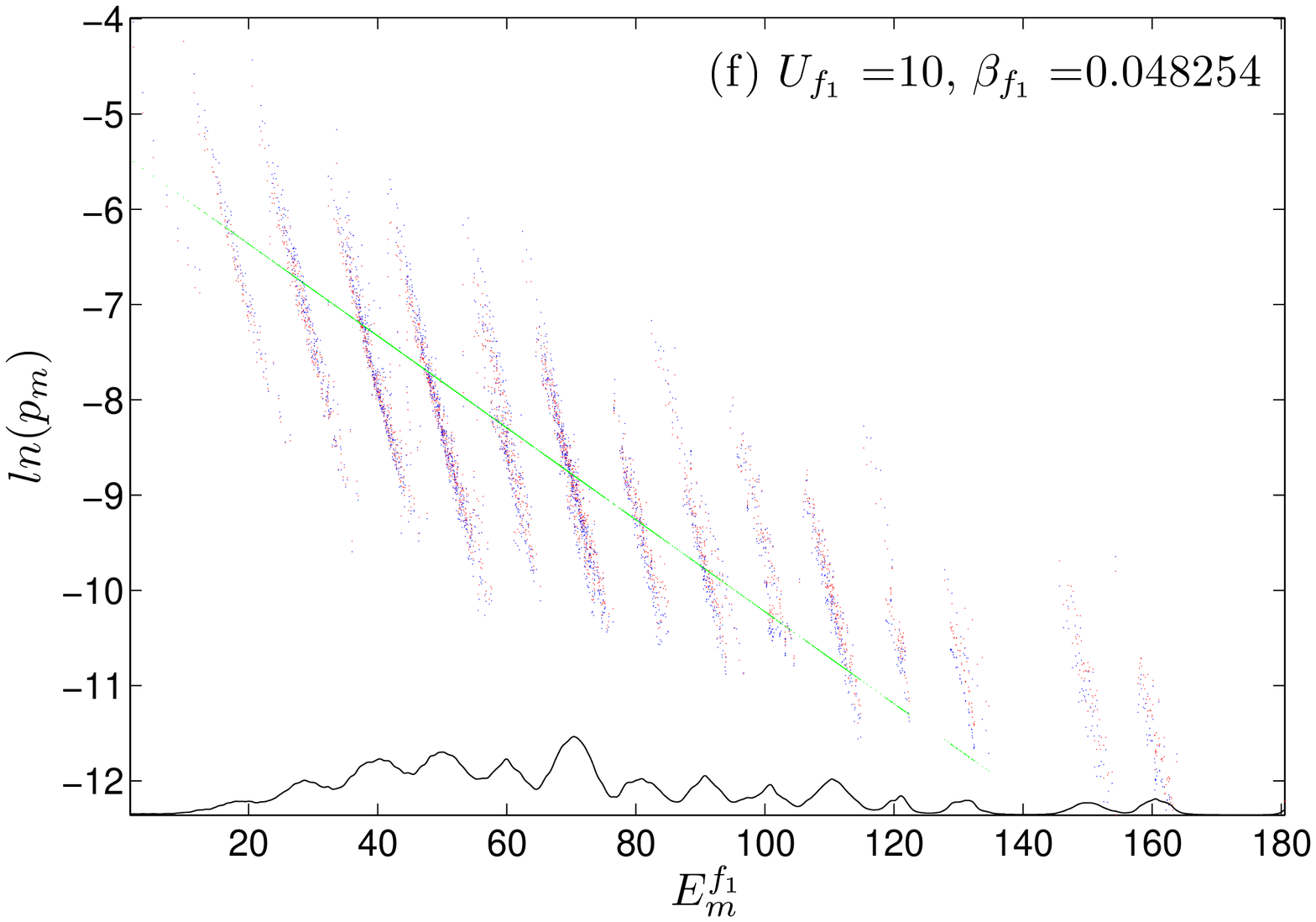}
\end{minipage}
\caption{(Color online) Semilog plots of $p_m$ versus the
eigenvalues $E^{f_1}_m$ (red dots). The initial state is
the same for all the figures, with parameters
$(M,N,q,D_q)=(9,9,1,2700)$, $U_i=1$, and $\beta_i=0.3$. The
quenched values of $U$ and the fitting inverse temperatures
$\beta_{f_1}$ are shown in the inserts. For comparison, the
data with $\rho_c$ (green dots) and $p'_m$ (blue dots) are
also shown. The black lines at the bottom depict the
coarse-grained density of states of $H_{f_1}$ (just for
reference, not corresponding to the vertical
axis).\label{fig1}}
\end{figure*}
\begin{figure*}[tb]
\begin{minipage}[b]{0.30 \textwidth}
\centering
\includegraphics[ width=\textwidth]{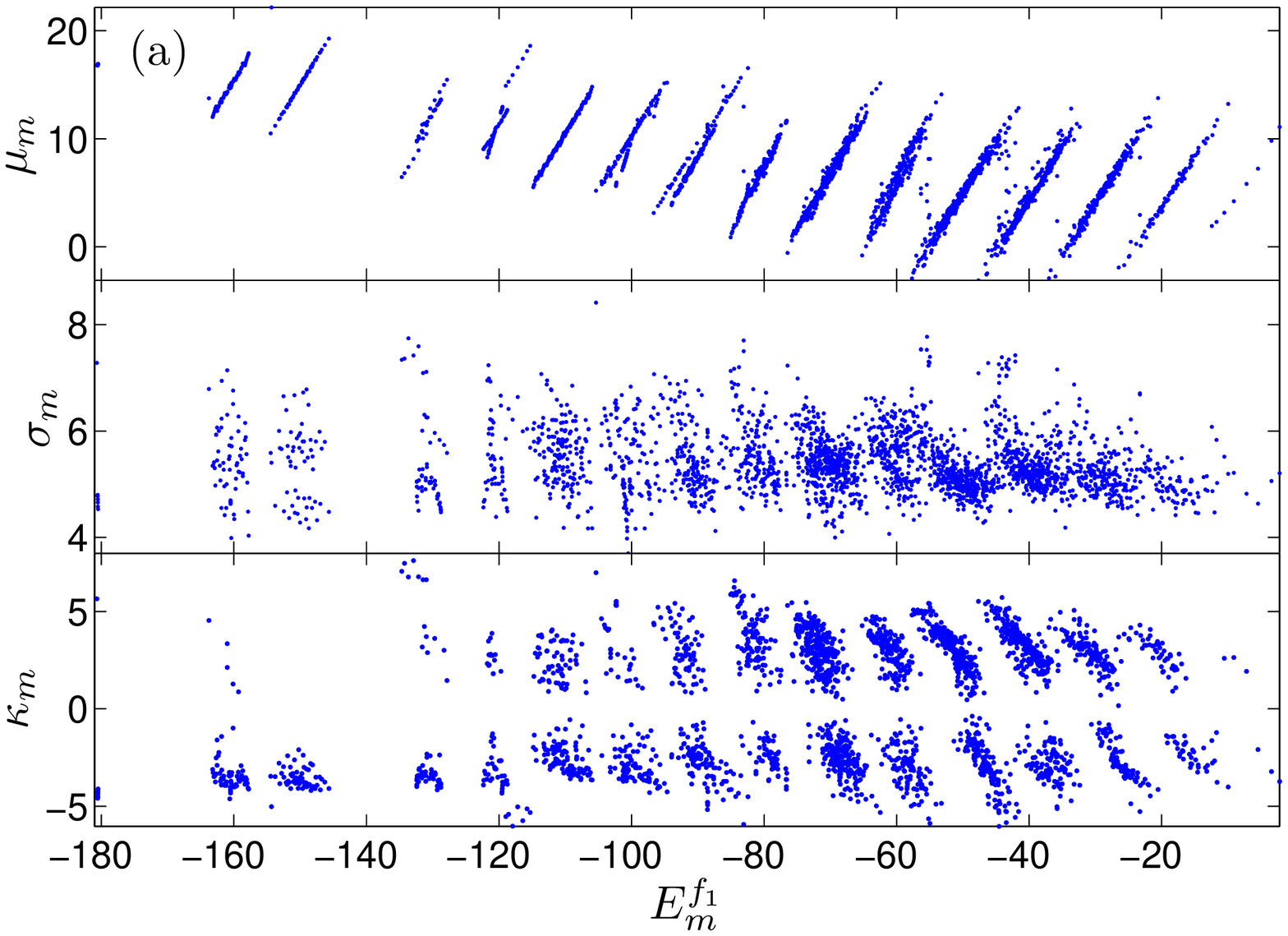}
\end{minipage}
\begin{minipage}[b]{0.30 \textwidth}
\centering
\includegraphics[ width=\textwidth]{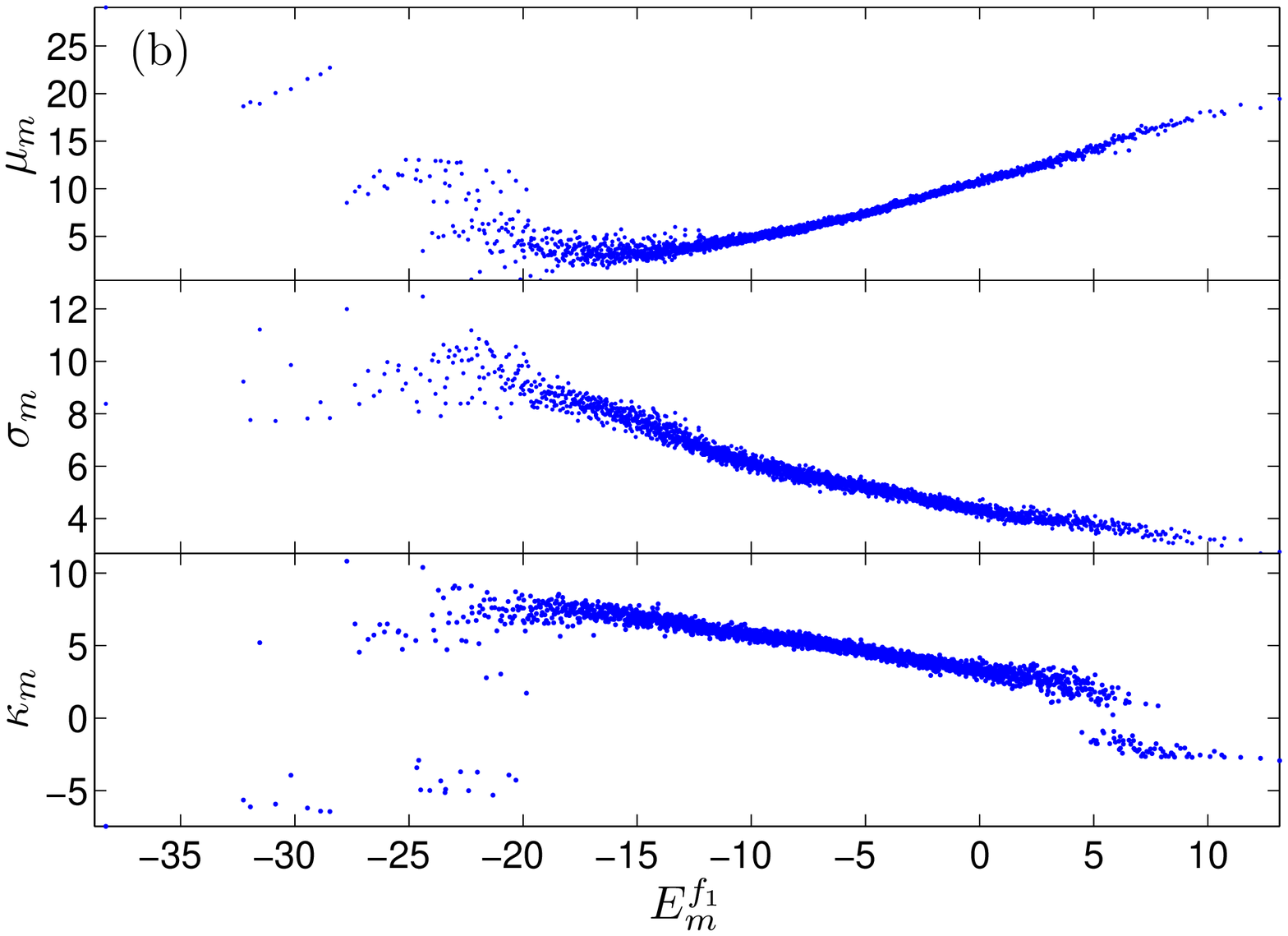}
\end{minipage}
\begin{minipage}[b]{0.30 \textwidth}
\centering
\includegraphics[ width=\textwidth]{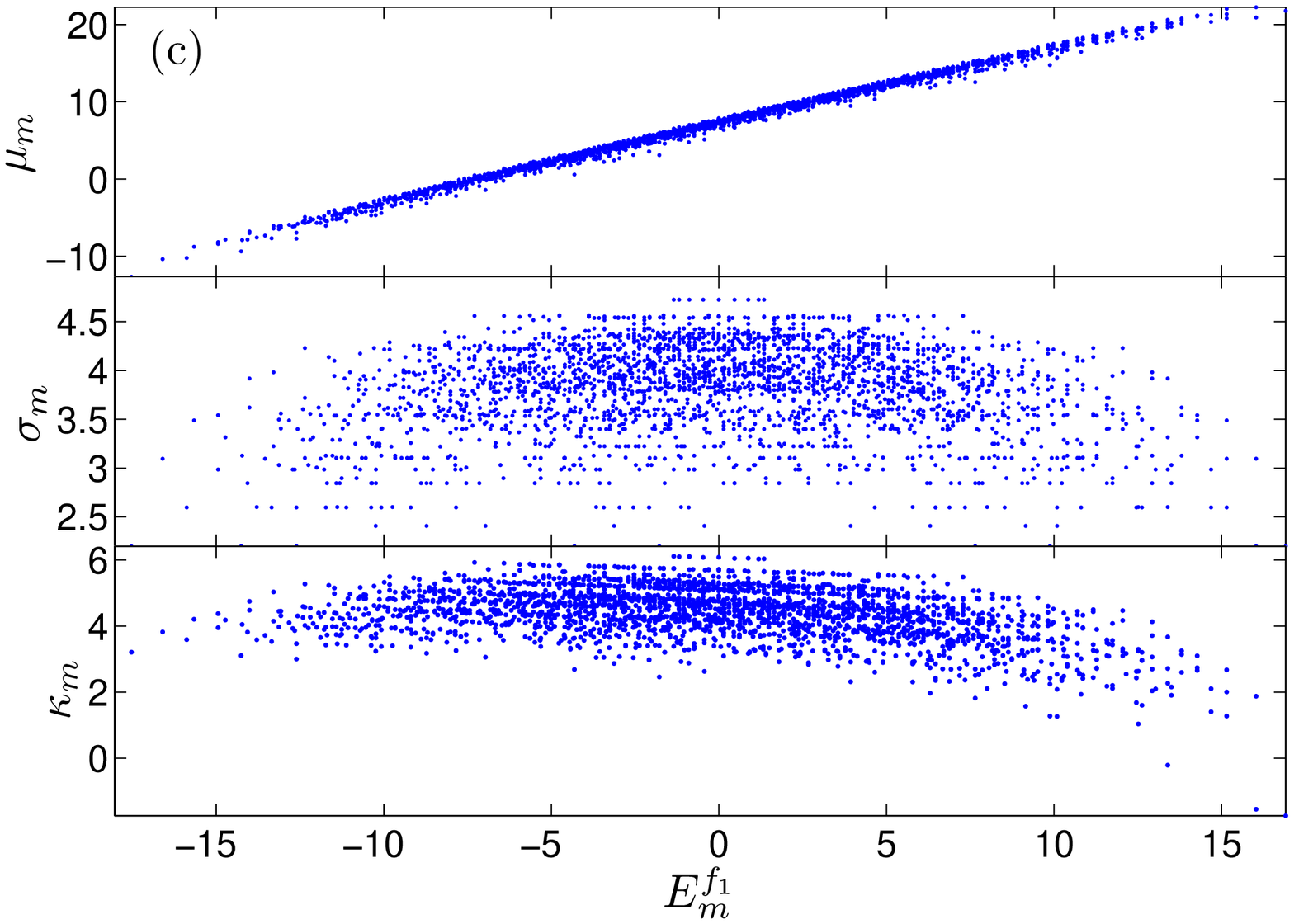}
\end{minipage}

\begin{minipage}[b]{0.30 \textwidth}
\centering
\includegraphics[ width=\textwidth]{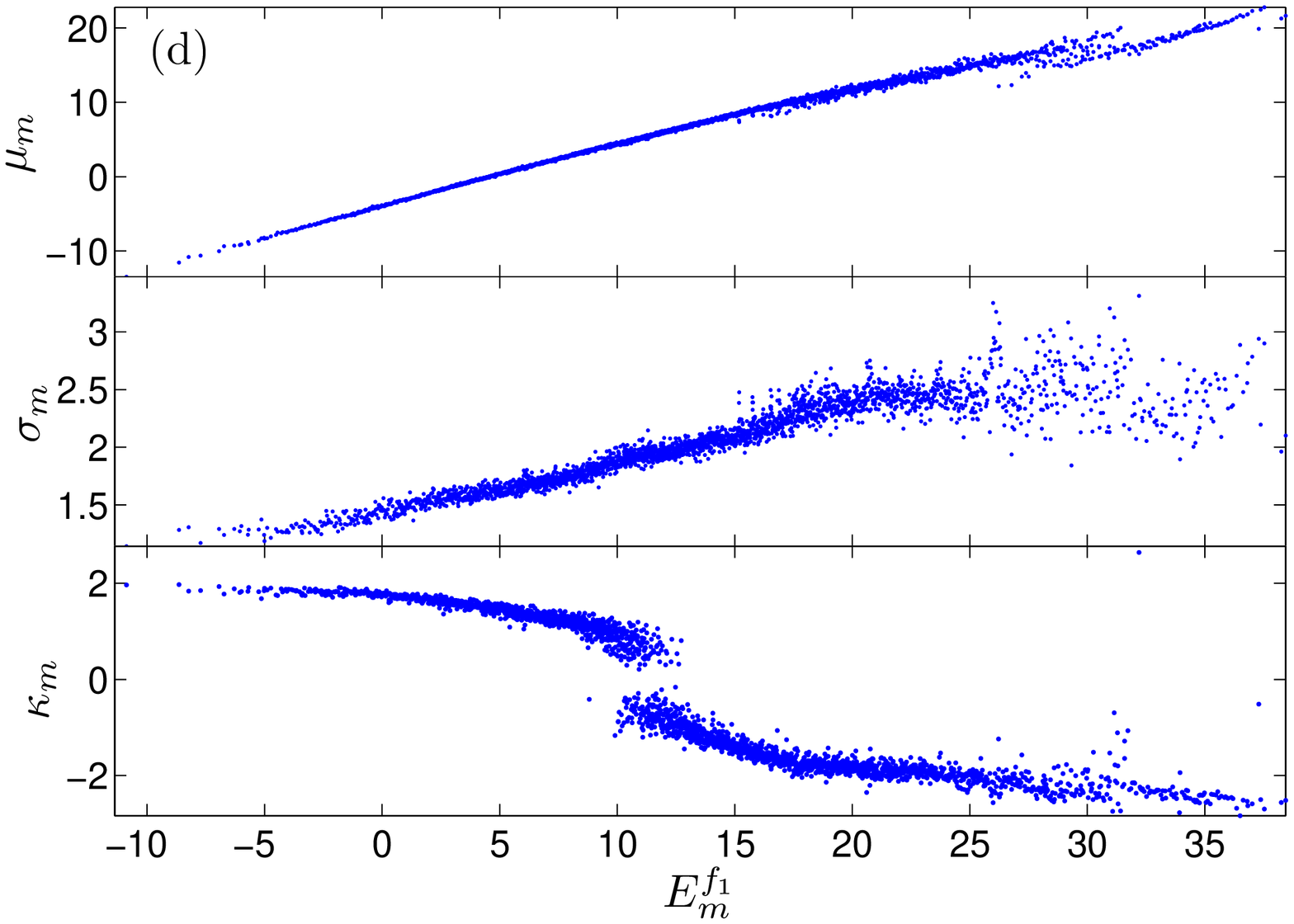}
\end{minipage}
\begin{minipage}[b]{0.30 \textwidth}
\centering
\includegraphics[ width=\textwidth]{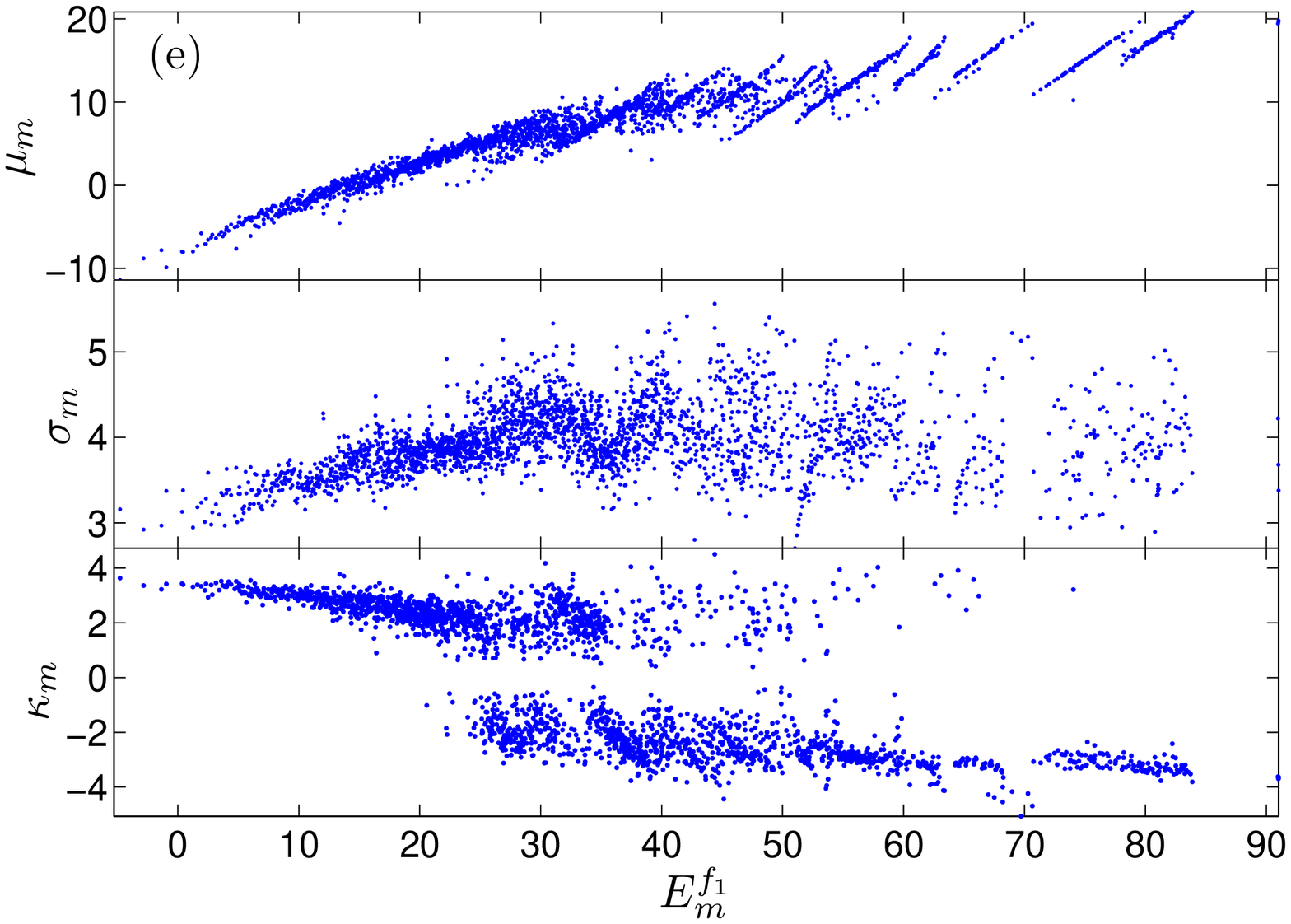}
\end{minipage}
\begin{minipage}[b]{0.30 \textwidth}
\centering
\includegraphics[ width=\textwidth]{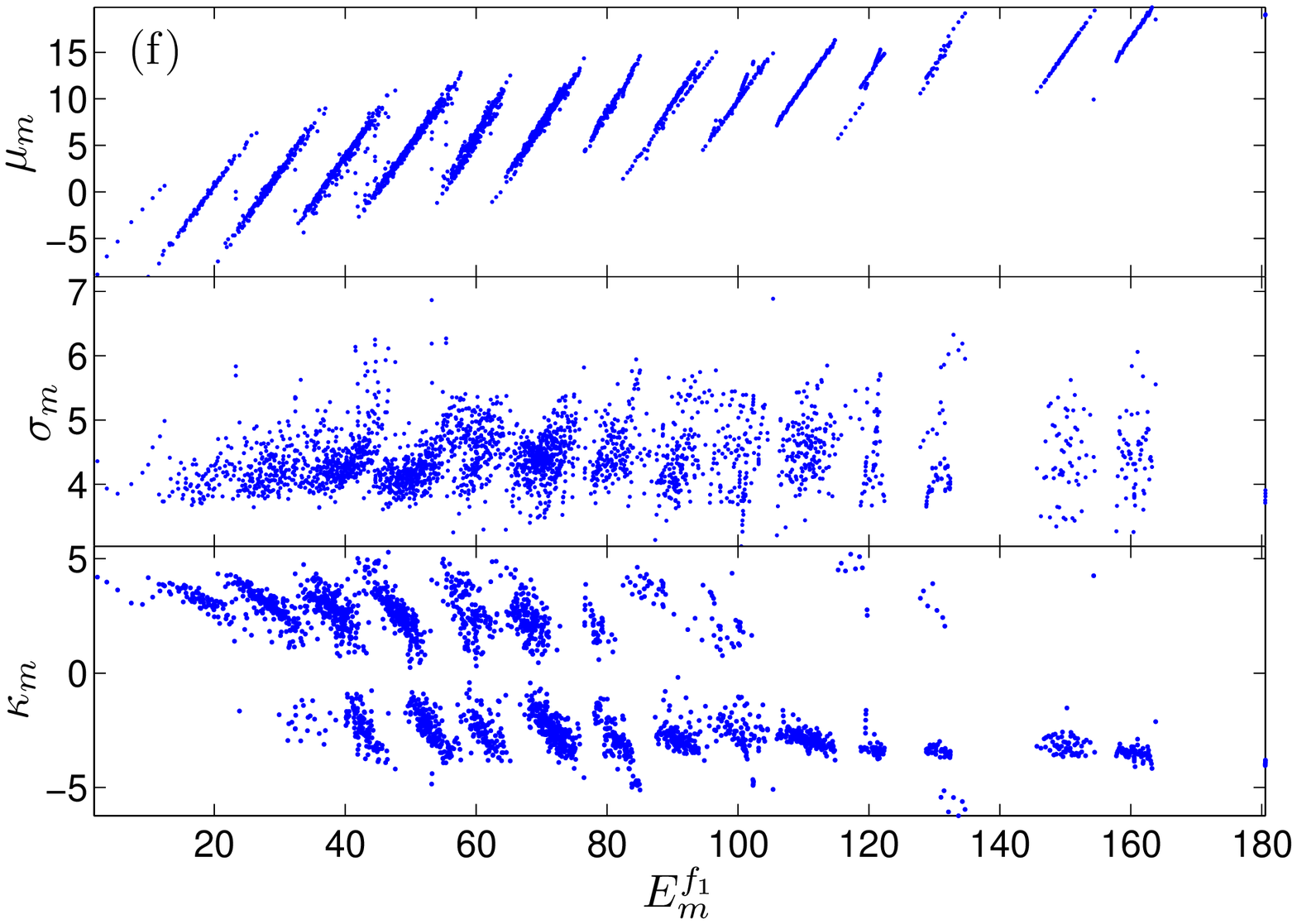}
\end{minipage}
\caption{(Color online) The parameters $\mu_m$, $\sigma_m$,
and $\kappa_m$ [see Eq.~(\ref{mean2})] characterizing the
probability distributions $P_m(E)$ associated with the
eigenstates of $H_{f_1}$. Note that
Figs.~\ref{fig2}a-\ref{fig2}f correspond to
Figs.~\ref{fig1}a-\ref{fig1}f, respectively. \label{fig2}}
\end{figure*}

Since the time-averaged density matrix provides an overall
information of the dynamics of the system, we look into it
first. In Fig.~\ref{fig1}, we consider the scenario of
starting from the same initial condition ($U_i=1$,
$\beta_i=0.3$) but quenching to six different values of
$U_{f_1}$ \cite{negative}. In each panel, the logarithms of
$p_m$ are plotted against the eigenvalues $E^{f_1}_m$ (red
dots). We have compared $\bar{\rho}$ with a canonical
ensemble density matrix $\rho_c$, which is defined as
\begin{equation}
\rho_c=\frac{e^{-\beta_{f_1}H_{f_1}}}{tr(e^{-\beta_{f_1} H_{f_1}} )}
\end{equation}
under the condition $tr(\rho_cH_{f_1})=tr(\bar{\rho}
H_{f_1})=tr(\rho_i H_{f_1})$. Here $\beta_{f_1}$, the final
inverse temperature, is the only fitting parameter. In
Fig.~\ref{fig1}, the green dots which form a straight line
correspond to $\rho_c$.

We see that $\bar{\rho}$ exhibits many interesting
features. In the case of $U_{f_1}=0$, $\bar{\rho}$ agrees
well with $\rho_c$ throughout the spectrum. In the case of
$U_{f_1}=2$, $\bar{\rho}$ agrees well with $\rho_c$ in the
lower part of the spectrum, while deviates from it
significantly in the higher part of the spectrum. But
overall the two are in good agreement since the weight of
the higher part is small. The case of $U_{f_1}=-1$ is
somewhat the reverse of the $U_{f_1}=2$ case. It is in the
lower part of the spectrum that $\ln p_m$ fluctuates
wildly. In the higher part $\ln p_m$ goes almost linearly.
Since the weight is dominated by the lower part, $\rho_c$
is not a good approximation of $\bar{\rho}$. In the strong
interaction limits of $U_{f_1}= \pm 10$, another feature
takes the place. As a whole the red dots do not fall close
to a single straight line, but they do form some stripes,
and the stripes are almost parallel with a common slope
close to $\beta_i$. It is easy to recognize that each
stripe corresponds to a bump in the density of states of
$H_{f_1}$.

In order to understand the various features in
Fig.~\ref{fig1}, we rewrite $p_m$ as
\begin{eqnarray} \label{pm2}
p_m = \frac{1}{Z_i} \int_{-\infty}^{+\infty} d E  e^{-\beta_i E} P_m(E) ,
\end{eqnarray}
where $P_m(E)=\sum_{n}  |\langle \psi_n^{i}  | \psi_m^{f_1}
\rangle |^2 \delta (E-E_n^i)$ is a probability distribution
\cite{prod} associated with $|\psi_m^{f_1} \rangle$. Note
that $P_m(E)$ is an intrinsic property of $| \psi_m^{f_1}
\rangle $ independent of $\beta_i$. We have tried to
characterize the distribution $P_m(E)$ by its mean $\mu_m$,
its second central moment $\sigma_m^2$, and its third
central moment $\kappa_m^3$, which are defined as follows,
\begin{subequations}\label{mean2}
\begin{eqnarray}
\mu_m &=& \int d E P_m(E) = \langle \psi^{f_1}_m | H_{i} | \psi^{f_1}_m \rangle , \label{ mean}\\
\sigma_m^2 &=& \int d E P_m(E) (E-\mu_m)^2, \label{sigma}\\
\kappa_m^3 &= &\int d E P_m(E) (E-\mu_m)^3. \label{kappa}
\end{eqnarray}
\end{subequations}
These quantities are presented in Fig.~\ref{fig2}. These
data enable us to understand Fig.~\ref{fig1}. Suppose for a
distribution $P_m(E)$ with $(\mu_m,\sigma_m)$, we define a
Gaussian distribution
\begin{equation}
P'_m(E)=\frac{1}{\sqrt{2\pi}\sigma_m} \exp\left(-\frac{(E-\mu_m)^2}{2\sigma_m^2} \right),
\end{equation}
which shares the same mean and variance with $P_m$ but has
vanishing third central moment. Replacing $P_m$ in
Eq.~(\ref{pm2}) by $P'_m$, we get an approximation of
$p_m$,
\begin{equation}\label{pmprime}
p'_m=\frac{1}{Z_i}\exp \left(-\beta_i \mu_m + \frac{1}{2} \beta_i^2 \sigma_m^2 \right).
\end{equation}
In Fig.~\ref{fig1}, $p'_m$ are represented by the blue
dots. We see that as a whole $p'_m$ is a good approximation
of $p_m$, except at the lower part of the spectrum in
Fig.~\ref{fig1}b. The reason is clear---the $\kappa_m$'s
there are the largest throughout all the figures, which
indicates that the corresponding distributions $P_m$ are
wide and asymmetric and thus cannot be well approximated
with a Gaussian distribution.

Now we can understand the good fittings in
Figs.~\ref{fig1}c and \ref{fig1}d. In these two cases,
$\mu_m$ is almost a linear function of $E^{f_1}_m$, and
$\sigma_m^2$ does not vary so much, therefore the exponent
in Eq.~(\ref{pmprime}) goes almost linearly with
$E^{f_1}_m$. The situation is similar in the higher part of
the spectrum in Fig.~\ref{fig2}b, and therefore we have a
good linear fitting for the higher spectrum part in
Fig.~\ref{fig1}b. In contrast, in Fig.~\ref{fig2}e, $\mu_m$
varies wildly for adjacent $E^{f_1}_m$, therefore we see in
Fig.~\ref{fig1}e large fluctuations about the straight
line. As for the parallel stripes in Figs.~\ref{fig1}a and
\ref{fig1}f, they are also understandable in terms of
Figs.~\ref{fig2}a and \ref{fig2}f, where $\mu_m$ form
parallel stripes. It is numerically checked and can be
argued that the slopes of the stripes are almost unity.
Actually we have
\begin{eqnarray}
E^{f_1}_m &=&\langle \psi^{f_1}_m |H_{f_1}| \psi^{f_1}_m \rangle \nonumber \\
&=& \langle \psi^{f_1}_m |H_{i}| \psi^{f_1}_m \rangle  +
(U_{f_1}-U_i) \langle \psi^{f_1}_m |H_{int}| \psi^{f_1}_m
\rangle,\quad \label{slope}
\end{eqnarray}
where $H_{int}= \frac{1}{2} \sum_{l=1}^M a_l^\dagger
a_l^\dagger a_l a_l$. Note that in the limit of large
$|U_{f_1}/J|$, the kinetic term in the Hamiltonian
(\ref{h}) can be viewed as a perturbation to the second
interaction term. The spectrum of the latter is highly
degenerate and consists of integral multipliers of
$U_{f_1}$. The effect of the perturbation is to mix up the
eigenstates of the interaction Hamiltonian with different
eigenvalues and smooth the spectrum. That is why there are
bumps in the density of states in Figs.~\ref{fig1}a and
\ref{fig1}f and two adjacent bumps are placed roughly
$U_{f_1}$ apart. By perturbation theory, it is easy to show
that the second term in Eq.~(\ref{slope}) varies on the
order of $J^2/|U_{f_1}| \ll J$ among eigenstates belonging
to the same bump. Therefore, approximately we have
$\mu_m=E^{f_1}_m-const$ for each bump and this explains why
the stripes in Figs.~\ref{fig2}a and \ref{fig2}f are of
slope unity. In turn it explains [with the help of
Eq.~(\ref{pmprime})] why we have the parallel stripes in
Figs.~\ref{fig1}a and \ref{fig1}f, and especially the
slopes are approximately $\beta_i$.

\begin{figure}[tbh]
\centering
\includegraphics[width=0.35\textwidth]{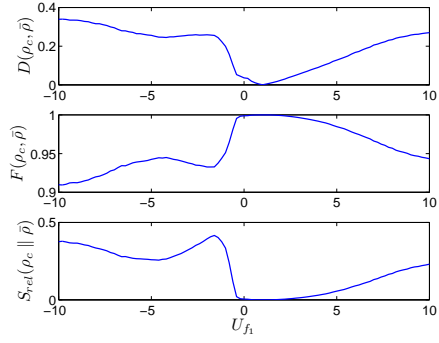}
\caption{\label{fig3} The distance $D$ and fidelity $F$
between $\rho_c$ and $\bar{\rho}$, and the relative entropy
of $\rho_c$ with respect to $\bar{\rho}$, as functions of
$U_{f_1}$. The initial state is the same as in
Fig.~\ref{fig1}.}
\end{figure}

It seems in Fig.~\ref{fig1} that $\rho_c$ is a good
approximation of $\bar{\rho}$ only when $|U_{f_1}-U_i|$ is
small. In Fig.~\ref{fig3}, we employ the tools of distance
$D$, fidelity $F$, and relative entropy $S_{rel}$ (for the
definitions see \cite{chuang}) between two density matrices
to quantify the difference or resemblance between $\rho_c$
and $\bar{\rho}$. There it is clear that only in the range
of $|U_{f_1}-U_i|\leq 1$, we have $(D,1-F,S_{rel})\ll 1$,
which means $\bar{\rho}$ is close to $\rho_c$. In the
subsequent subsection we will see that only in this range
the expectation values of some generic physical observables
according to $\bar{\rho}$ and $\rho_c$ agree well.

\subsection{Time evolution}
\begin{figure*}[tb]
\begin{minipage}[b]{0.30 \textwidth}
\centering
\includegraphics[ width=\textwidth]{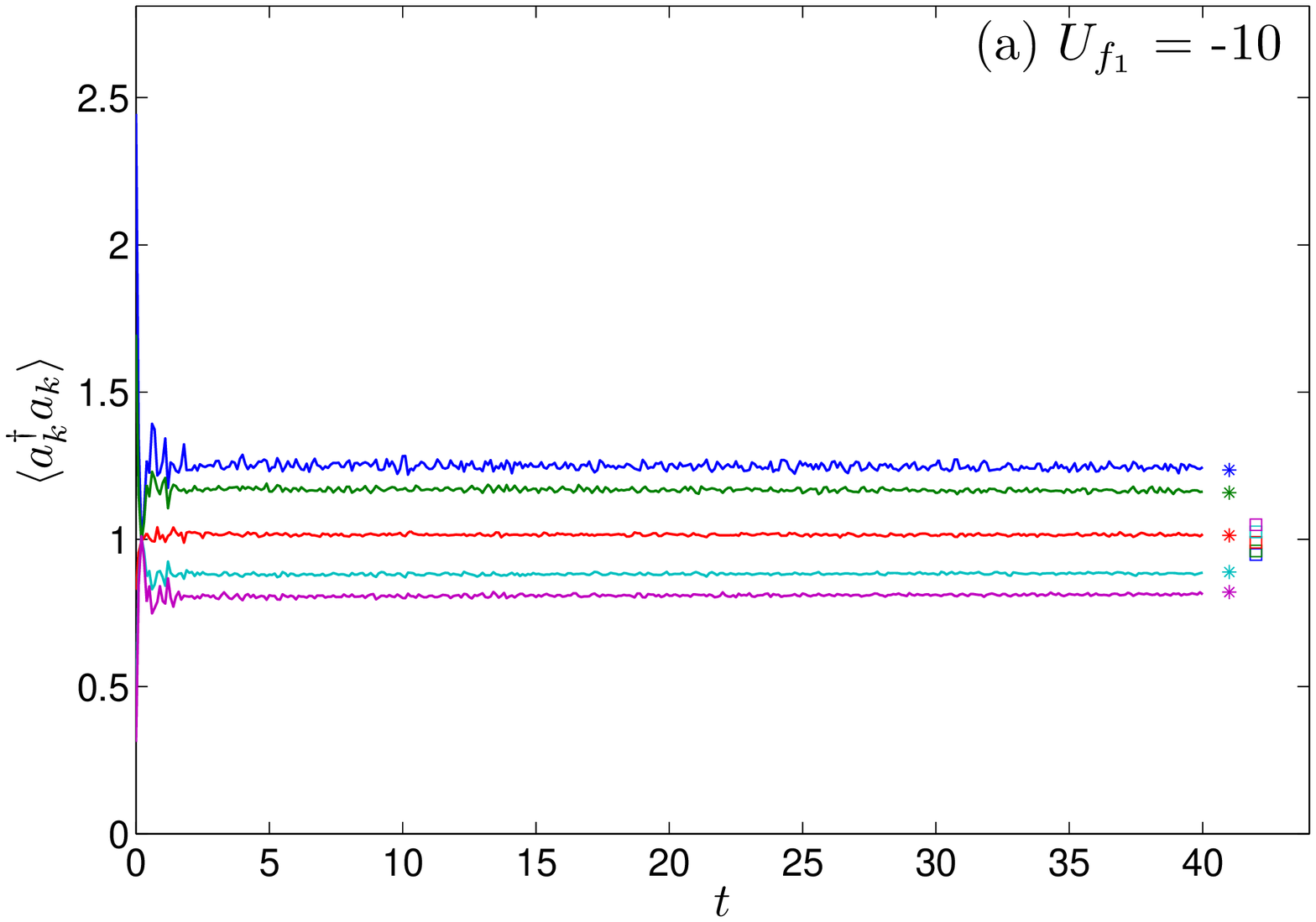}
\end{minipage}
\begin{minipage}[b]{0.30 \textwidth}
\centering
\includegraphics[ width=\textwidth]{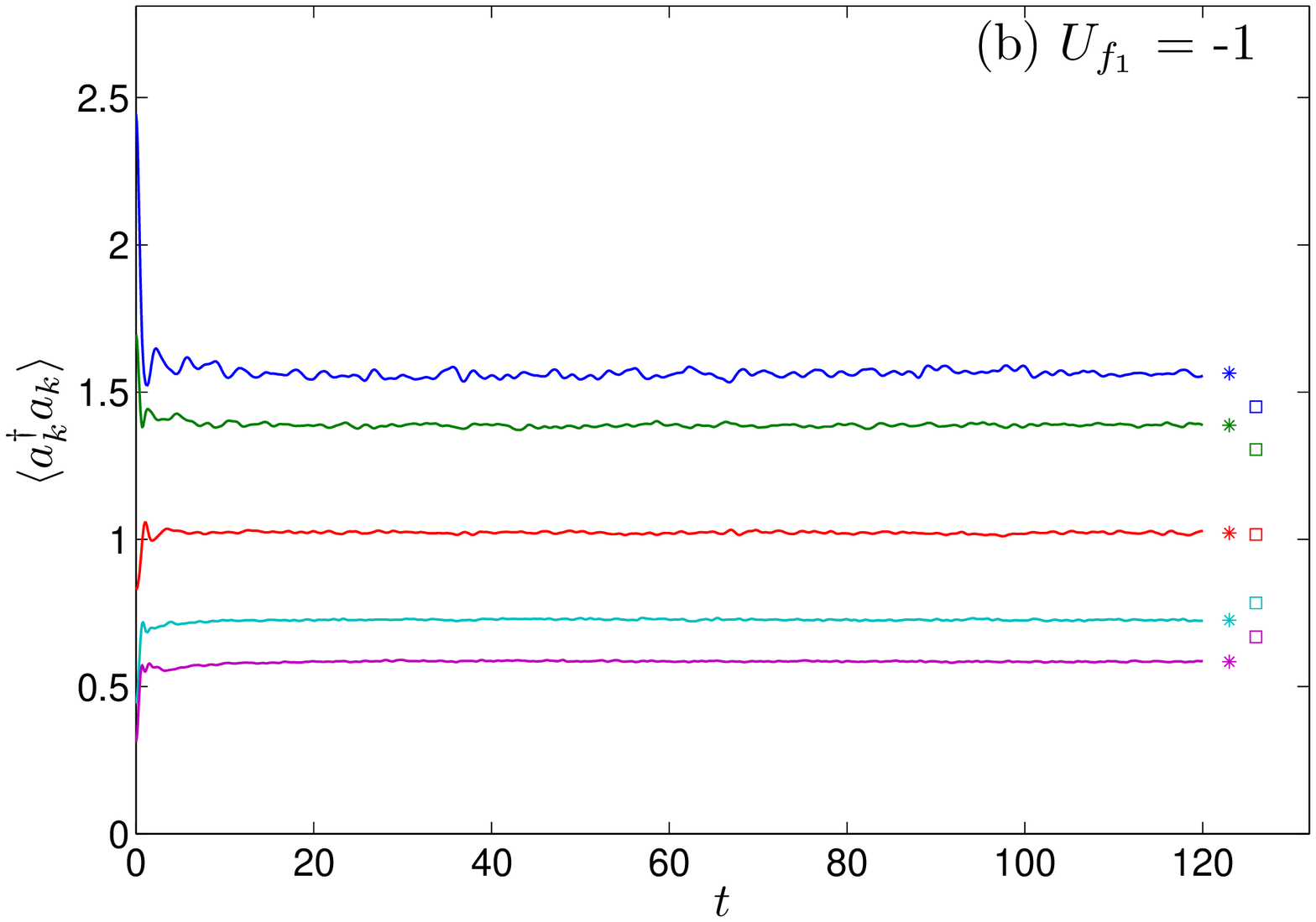}
\end{minipage}
\begin{minipage}[b]{0.30 \textwidth}
\centering
\includegraphics[ width=\textwidth]{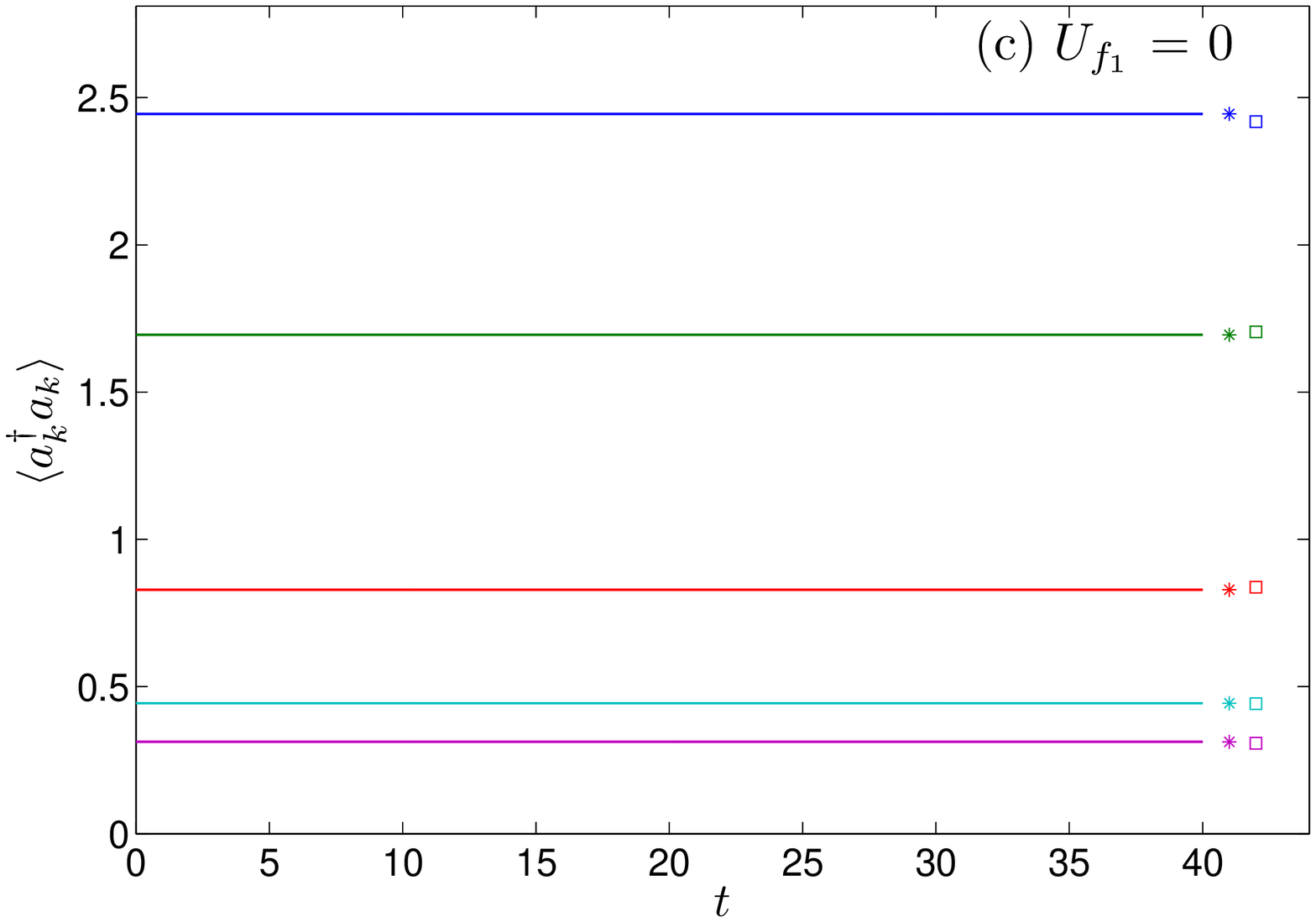}
\end{minipage}

\begin{minipage}[b]{0.30 \textwidth}
\centering
\includegraphics[ width=\textwidth]{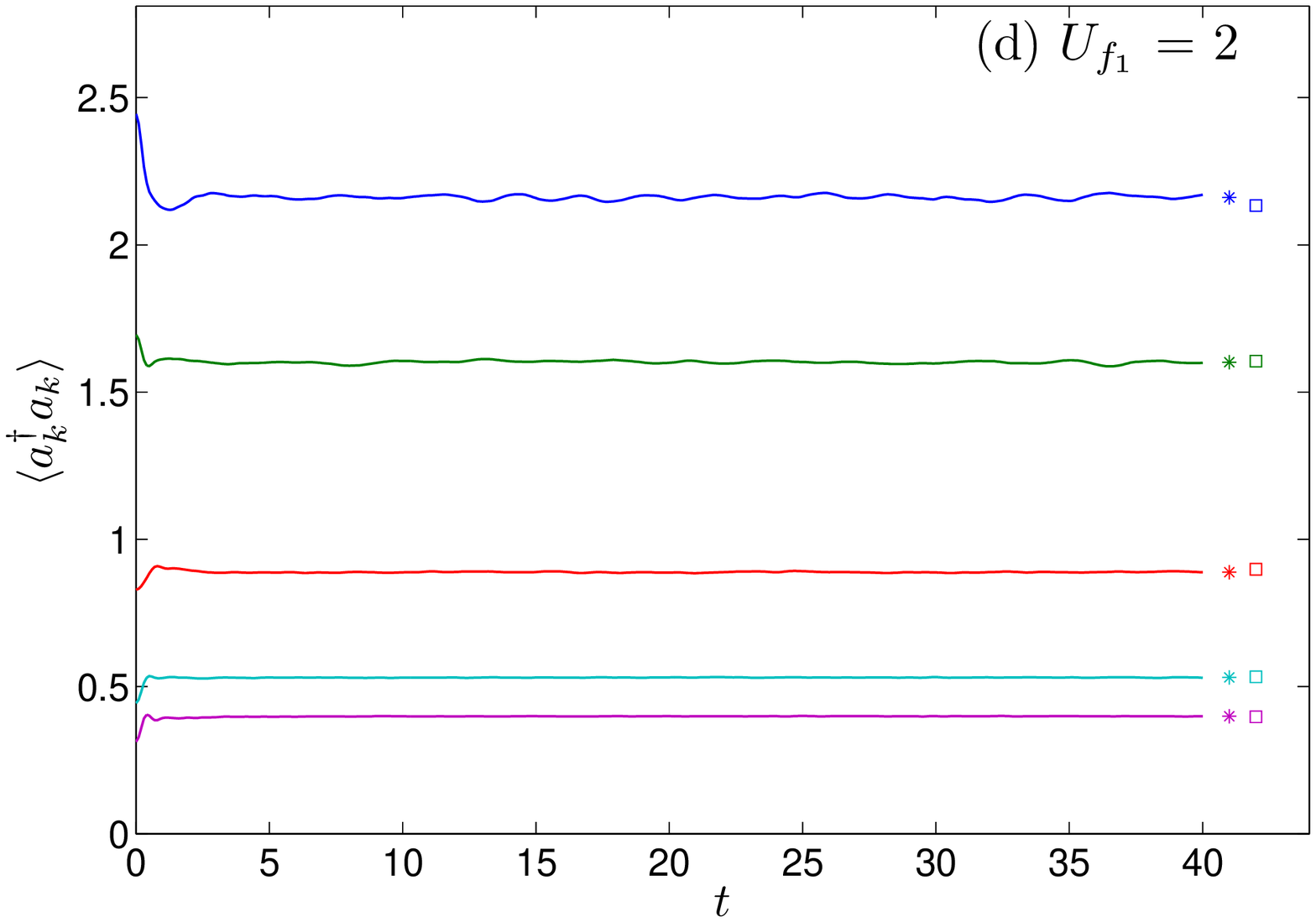}
\end{minipage}
\begin{minipage}[b]{0.30 \textwidth}
\centering
\includegraphics[ width=\textwidth]{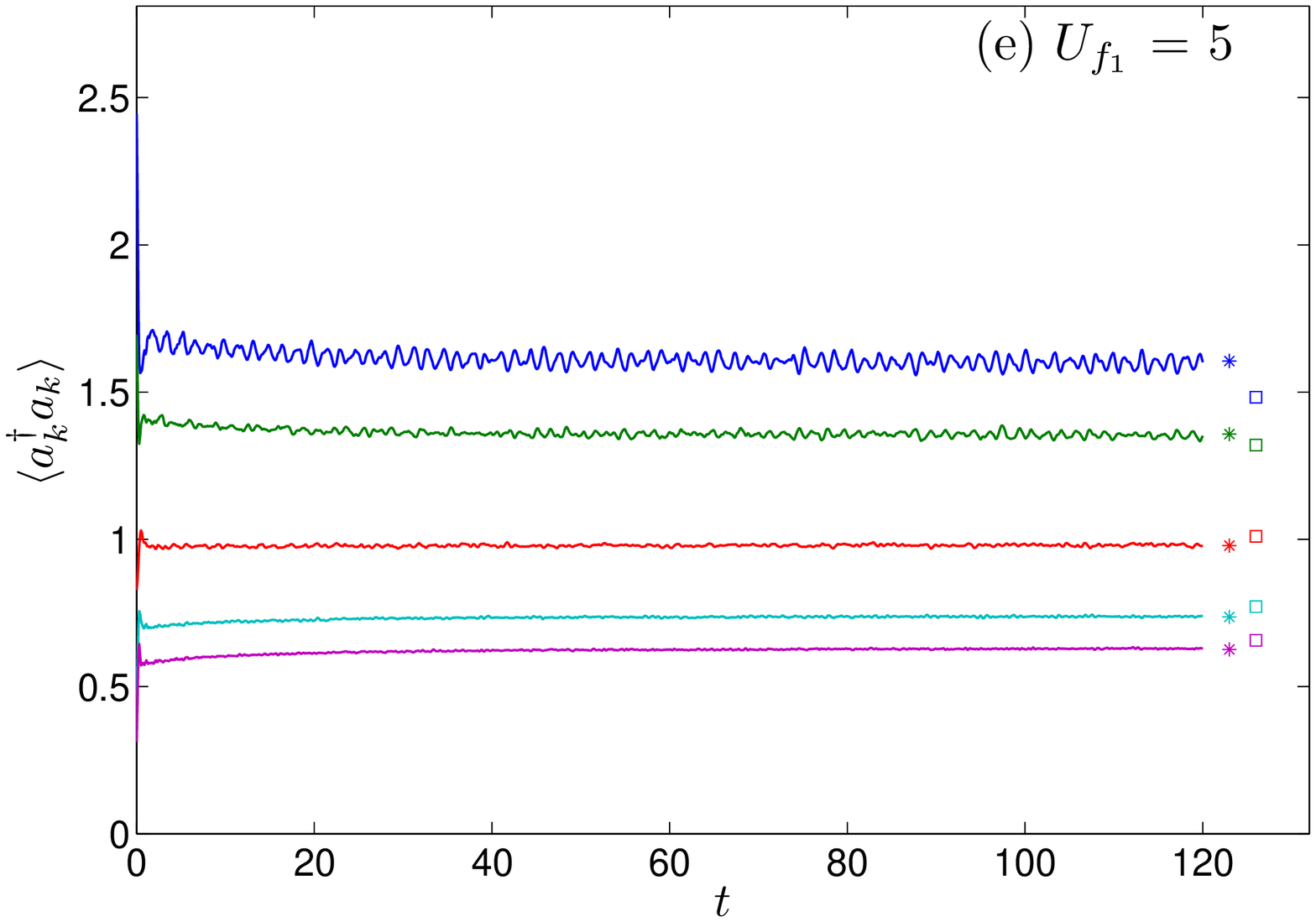}
\end{minipage}
\begin{minipage}[b]{0.30 \textwidth}
\centering
\includegraphics[ width=\textwidth]{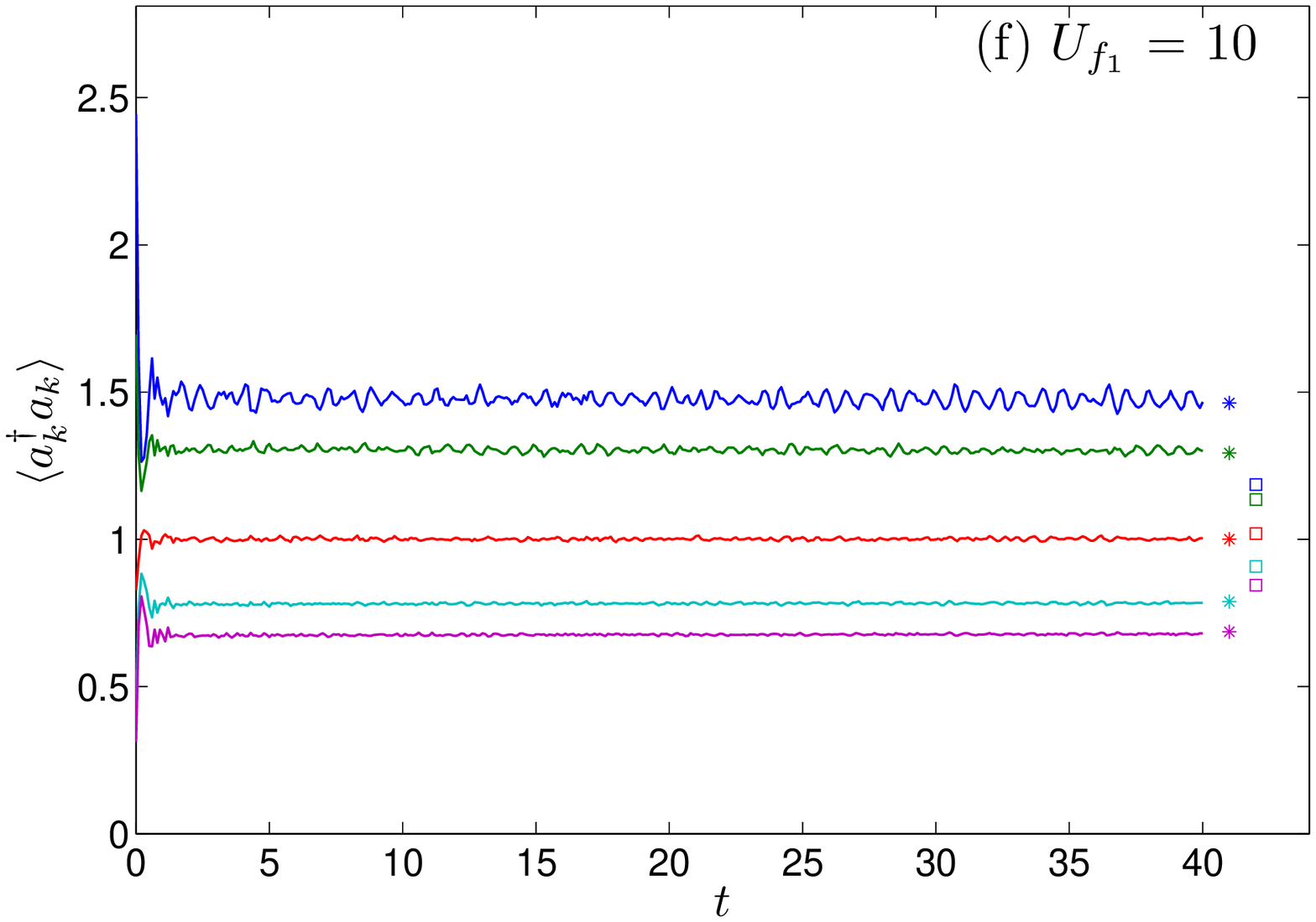}
\end{minipage}
\caption{(Color online) Time evolution of the populations
on the Bloch states $\langle a^\dagger_k a_k \rangle $. The
figures correspond to those in Fig.~\ref{fig1} in a
one-to-one manner. In each figure, from up to down, the
five lines correspond to $k=0,\ldots,4$. Another $k$'s are
now shown because $\langle a^\dagger_k a_k \rangle$ and
$\langle a^\dagger_{M-k} a_{M-k} \rangle$ are close to each
other all the time. For each line, the markers of the same
color on the right hand side indicate the average value
predicted by $\bar{\rho}$ ($*$) or value predicted by
$\rho_c$ ($\square$), respectively. Note that in (b) and
(e), the time span investigated is longer than that in
others. This is because the transient times in (b) and (e)
are relatively longer. \label{fig4}}
\end{figure*}

We now proceed to study the time evolution of the system
after the quench. In Fig.~\ref{fig2}, we show the time
evolution of the populations on the Bloch states $\langle
a^\dagger_k a_k \rangle$. The six sub-figures correspond to
those in Fig.~\ref{fig1} respectively. For all the
$U_{f_1}$'s and all the $k$'s, $\langle a_k^\dagger a_k
\rangle$ equilibrate to their average values after a
transient time, which is relatively longer in the cases of
$U_{f_1}=-1$ and $5$. In the special case of $U_{f_1}=0$,
there is no fluctuation at all. The reason is simply that
in this case, $a^\dagger_k a_k$ are conserved. We see that
the time-averaged values of $\langle a_k^\dagger a_k
\rangle $ predicted by $\bar{\rho}$ ($*$) and $\rho_c$
($\square$) agree relatively well in the cases of
$U_{f_1}=0$ and $2$. This is consistent with the closeness
between $\bar{\rho}$ and $\rho_c$ for these two values of
$U_{f_1}$, as revealed in Fig.~\ref{fig1} and
Fig.~\ref{fig3}. Here we would say the system thermalizes
well in the $U_{f_1}=2$ case, however, we would refrain
making the same statement for the $U_{f_1}=0$ case. The
reason will be clear in the next Section.

Figure \ref{fig4} is about a finite-sized system with some
specific initial condition. However, here we have some
general statements. We argue that in the thermodynamic
limit ($M,N\rightarrow \infty$ with $N/M$ fixed), as long
as initially the system is at finite-temperature thermal
equilibrium and described by a canonical ensemble density
matrix as (\ref{rhoi}), we should see steady behaviors of
the physical variables like $a^\dagger_k a_k$.

Let $A=a^\dagger_k a_k$ and let $A=\sum_{mn} A_{mn}
|\psi^{f_1}_m\rangle \langle \psi^{f_1}_n |$ in the
representation of $\{|\psi^{f_1}_m \rangle \}$. The
ensemble-averaged value of $A$ at time $t$ is
\begin{equation}
a(t)=\sum_{mn} \rho_{mn} A_{nm} \exp[-i(E^{f_1}_m-E^{f_1}_n)t],
\end{equation}
where $\rho_{mn}\equiv \langle \psi^{f_1}_m | \rho_i |
\psi^{f_1}_n \rangle$. Its time-averaged value is
\begin{equation}
\bar{a}=\lim_{T\rightarrow \infty}\frac{1}{T}\int_0^T dt a(t)=\sum_{m} \rho_{mm} A_{mm}.
\end{equation}
Here note that for a generic Hamiltonian $H_{f_1}$, there
is no level degeneracy. The time-averaged value of $a^2(t)$
is \cite{peter}
\begin{eqnarray}
\overline{a^2}&=&\lim_{T\rightarrow \infty}\frac{1}{T}\int_0^T dt a^2(t) \nonumber \\
&=& \sum_{mp}  \rho_{mm} A_{mm} \rho_{pp} A_{pp} +\sum_{m\neq n} \rho_{mn} A_{nm} \rho_{nm} A_{mn}  \nonumber \\
&=& \sum_{m} \rho_{mm} A_{mm}\sum_p \rho_{pp} A_{pp} +\sum_{m\neq n} |\rho_{mn}|^2 |A_{mn}|^2 \nonumber \\
&=& \bar{a}^2 + \sum_{m\neq n} |\rho_{mn}|^2 |A_{mn}|^2.
\end{eqnarray}
Note that here it is assumed that there is no degeneracy of
energy gaps. Thus we have for the variance of $a(t)$ in
time, $\Delta^2 a= \overline{a^2}-\bar{a}^2$,
\begin{equation}
\Delta^2 a=\sum_{m\neq n} |\rho_{mn}|^2 |A_{mn}|^2\leq \sum_{m n} |\rho_{mn}|^2 |A_{mn}|^2.
\end{equation}
Since $A$ is  semi-positive definite and bounded, we have
$|A_{mn}|^2\leq A_{mm} A_{nn} \leq N^2$. Thus we have
\begin{equation}\label{bound1}
\Delta^2 a\leq N^2 \sum_{mn} |\rho_{mn}|^2.
\end{equation}
Here we note that the summation is the square of the
Frobenius norm of $\rho_i$ in the representation of
$\{|\psi^{f_1}_m\rangle \}$, which is invariant in all
representations and is preserved by an arbitrary unitary
evolution \cite{horn}. Explicitly, we have
\begin{eqnarray}
\sum_{mn} |\rho_{mn}|^2&=&\sum_{mn} \langle \psi^{f_1}_m |\rho_i | \psi^{f_1}_n \rangle \langle \psi^{f_1}_n | \rho_i | \psi^{f_1}_m \rangle \nonumber \\
&=& \sum_{m} \langle \psi^{f_1}_m | \rho_i^2 | \psi^{f_1}_m \rangle = \sum_{m} \langle \psi^i_m | \rho_i^2 |\psi^i_m \rangle \nonumber \\
&=& \sum_m (p_m^i)^2. \label{fnorm}
\end{eqnarray}

We argue that this quantity, which depends only on the
initial state, decays exponentially with the size $M$. Let
$E^i_m$ increase with $m$. We have
\begin{eqnarray} \label{bound2}
\sum_m (p_m^i)^2 < p_1^i &=&\frac{e^{-\beta_i
E^i_1}}{Z_i}=\frac{e^{-\beta_i E^i_1}}{e^{-\beta_i
F_i}}\simeq \frac{e^{-\beta_i \alpha M}}{e^{-\beta_i \gamma
M}},\quad
\end{eqnarray}
as $M\rightarrow  \infty$. Here in the $\simeq$ relation we
used the fact the ground state energy $E^i_1$ of $H_i$
scales linearly with $M$ and so does the free energy $F_i$
of the initial state \cite{linear}. The coefficients
$\alpha$ and $\gamma$ are independent of $M$. Moreover, it
is easy to see that $\alpha \geq \gamma$ for any $\beta_i$,
with the equality taken only in the limit of
$\beta_i=+\infty$ or $T_i=0^+$, and $\alpha-\gamma$
increases monotonically with $T_i$. This makes sure that
$p_1^i$ would not grow exponentially with $M$ and transcend
unity.

With (\ref{bound1}) and (\ref{bound2}), we get an upper
bound for $\Delta a$,
\begin{equation} \label{bound3}
{\Delta a} \leq c M \exp(-\beta_i \theta M) ,\quad \theta =\frac{1}{2}(\alpha-\gamma ) \geq 0,
\end{equation}
where $c$ is some constant. The upper bound of $\Delta a$
helps us determine an upper bound for the probability of
finding $a(t)$ deviating away from the mean $\bar{a}$ by a
distance larger than $\epsilon$. Actually, following
Reimann \cite{peter}, using the Chebyshev inequality
\cite{feller}, we have
\begin{equation}
Prob(|a(t)-\bar{a}|>\epsilon)< \frac{\Delta^2 a}{\epsilon^2}.
\end{equation}
For a fixed value of $\epsilon$, the upper bound decreases
exponentially with the size of the system according to
(\ref{bound3}). It then follows the statement above.

Here some comments are worthy. Though in the derivation
above we have in mind a sudden quench, it is easy to see
that the conclusion actually applies to any type of quench
(e.g., the Hamiltonian can be changed continuously over
some period, as in \cite{sengupta,dimer}, or quenched
multiple times as in Sec.~\ref{sec4} below), as long as
after some point the Hamiltonian is never changed again.
The reason lies in that the Frobenius norm of the density
matrix $\rho(t)$ is conserved under unitary evolutions, and
thus is independent of the historical or the final values
of $H(t)$, but is determined entirely by the initial state.
As for the operator $A$, only the properties of
semi-positive-definiteness and boundedness are used. Thus
similar conclusions can apply to other operators such as
$a_k^\dagger a_k^\dagger a_k a_k$ and $a_l^\dagger
a_l^\dagger a_l a_l$, or operators in other models.
Finally, it should be mentioned that the conclusion relies
on the fact that the quantity in Eq.~(\ref{fnorm}) is
bounded by some exponentially decreasing function, which is
the case only at finite temperatures ($\beta_i < \infty$).
At zero temperature, the quantity in Eq.~(\ref{fnorm}) is
always equal to unity and thus the problem is still open.

\section{a second quench: typicality}\label{sec4}

\begin{figure*}[tb]
\begin{minipage}[b]{0.24 \textwidth}
\centering
\includegraphics[ width=\textwidth]{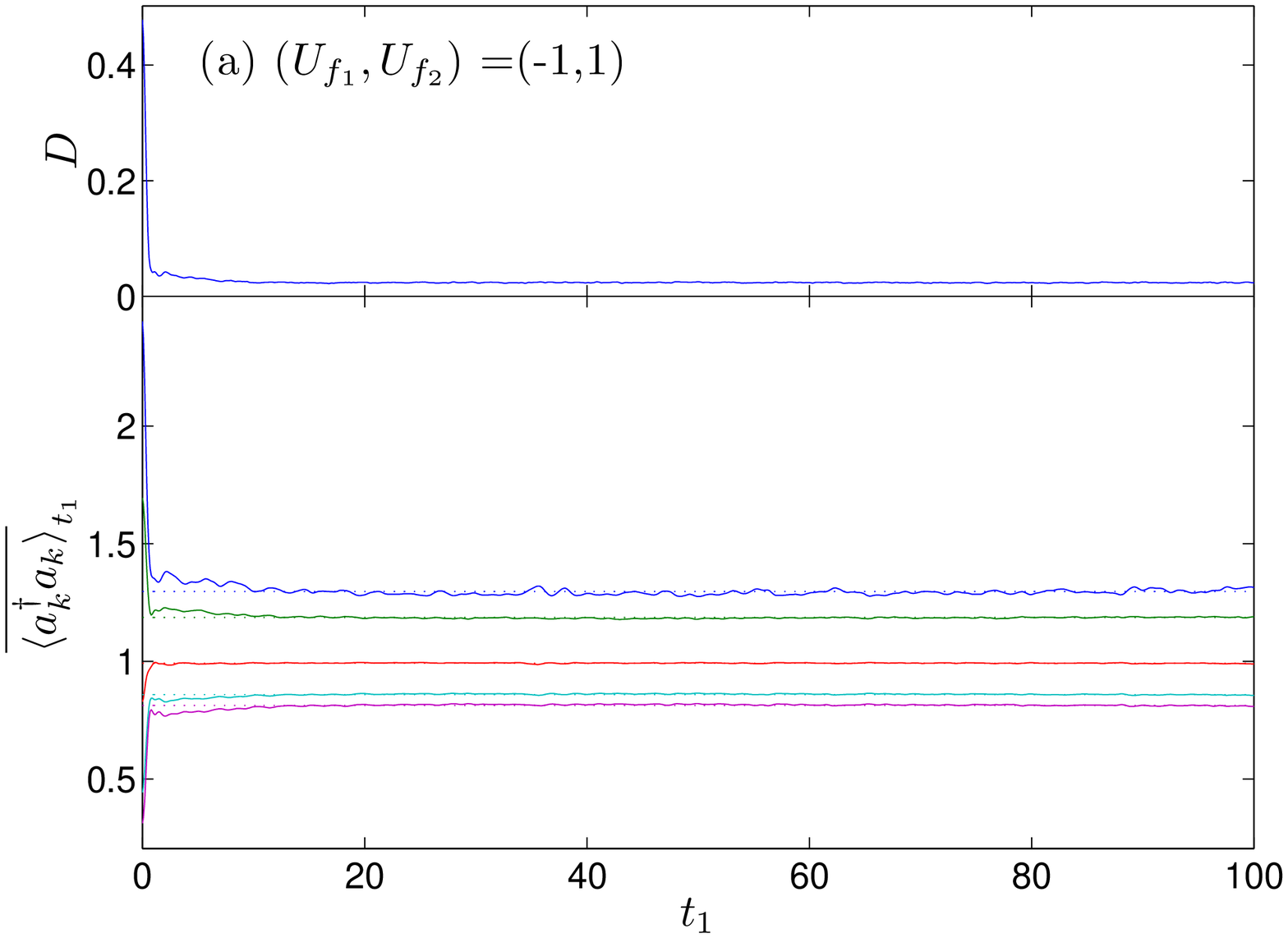}
\end{minipage}
\begin{minipage}[b]{0.24 \textwidth}
\centering
\includegraphics[ width=\textwidth]{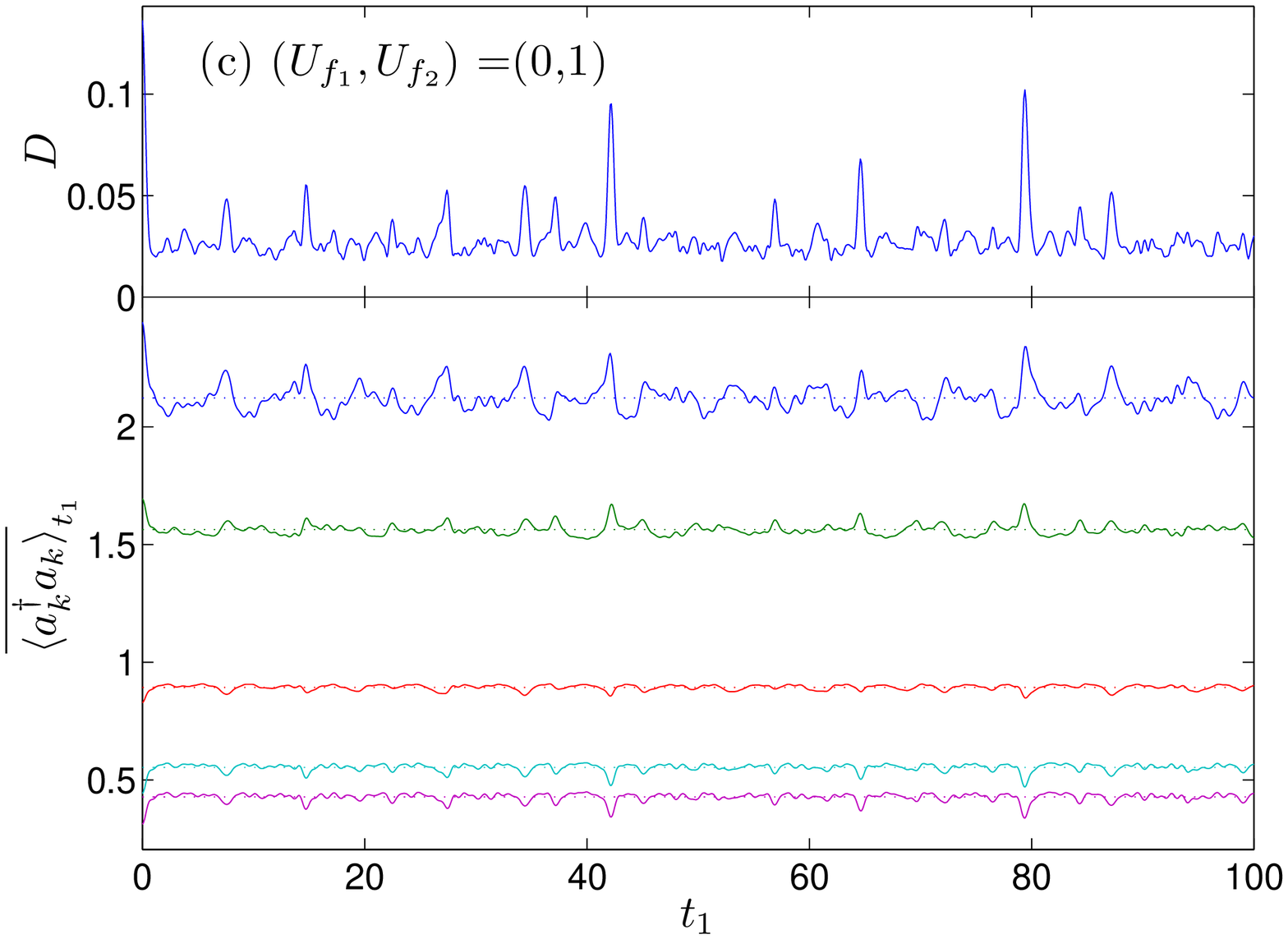}
\end{minipage}
\begin{minipage}[b]{0.24 \textwidth}
\centering
\includegraphics[ width=\textwidth]{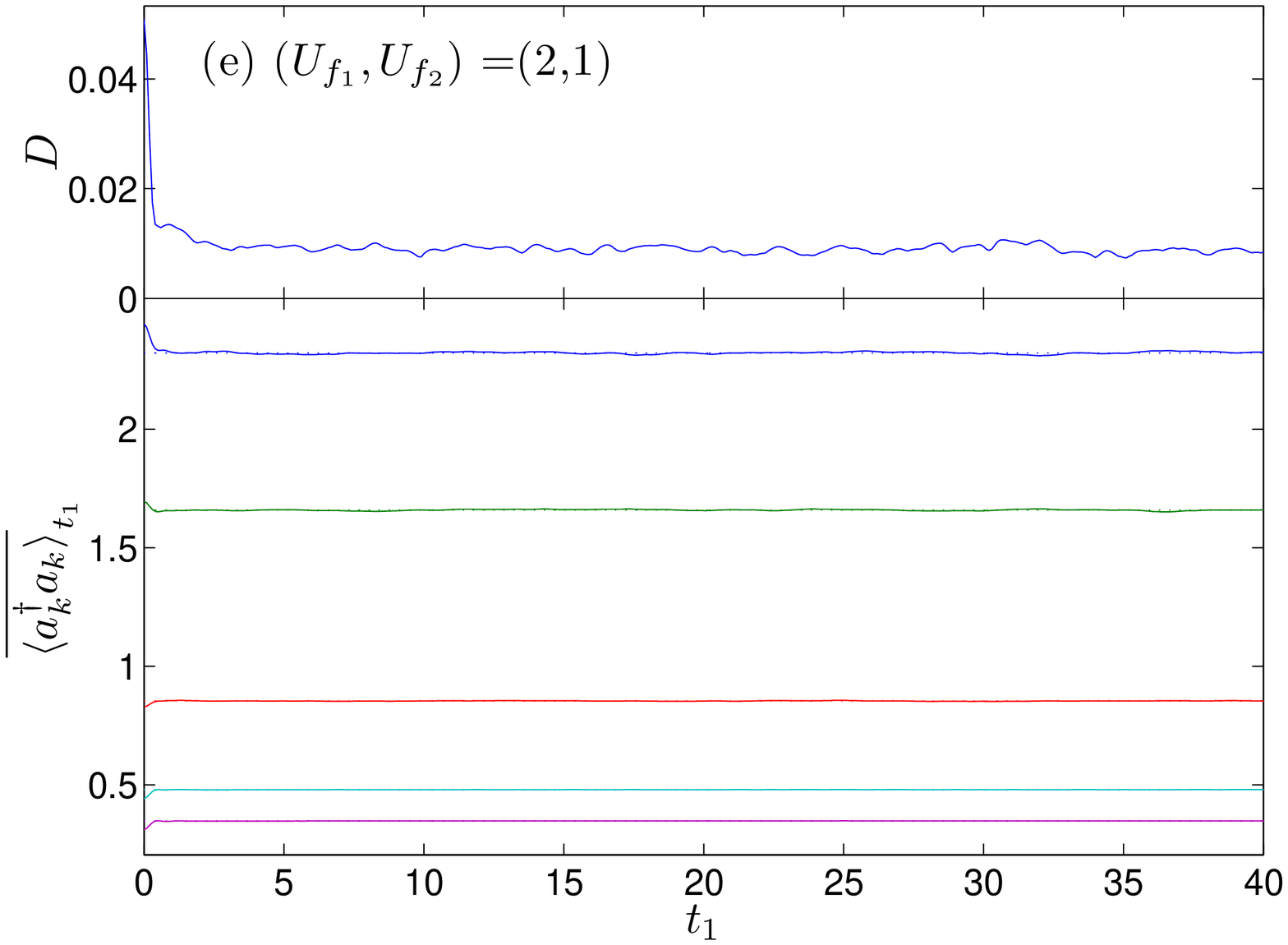}
\end{minipage}
\begin{minipage}[b]{0.24 \textwidth}
\centering
\includegraphics[ width=\textwidth]{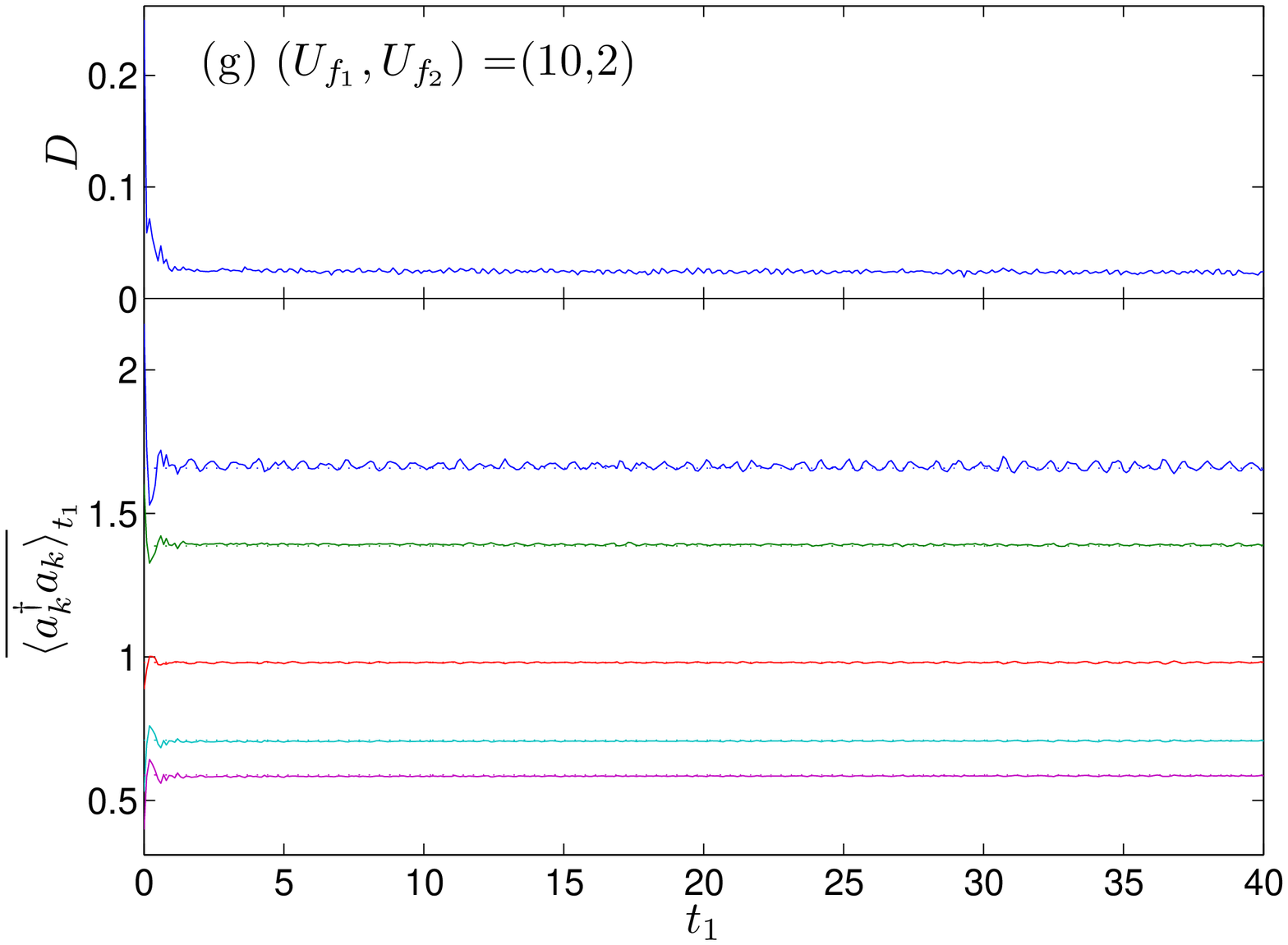}
\end{minipage}

\begin{minipage}[b]{0.24 \textwidth}
\centering
\includegraphics[ width=\textwidth]{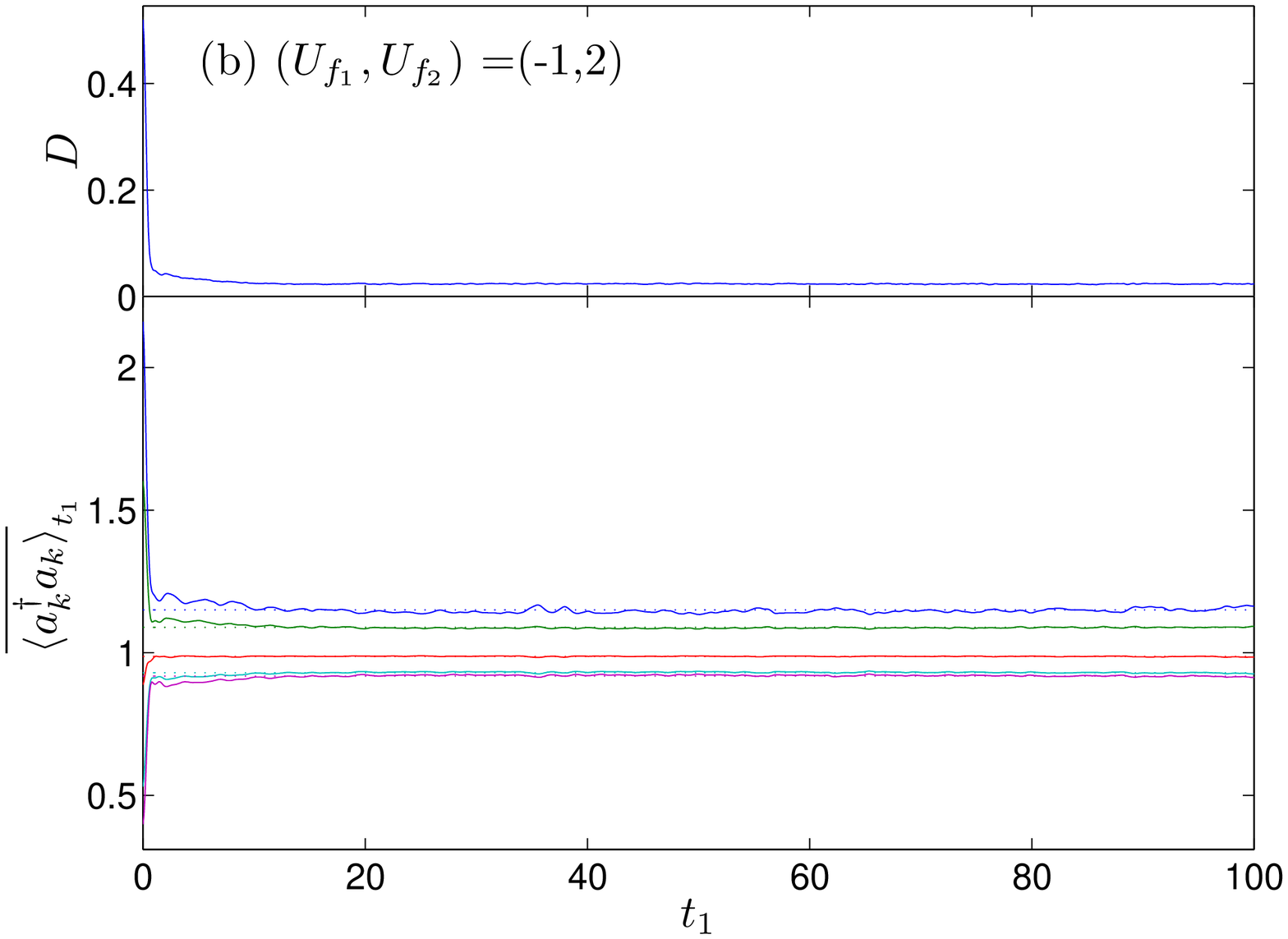}
\end{minipage}
\begin{minipage}[b]{0.24 \textwidth}
\centering
\includegraphics[ width=\textwidth]{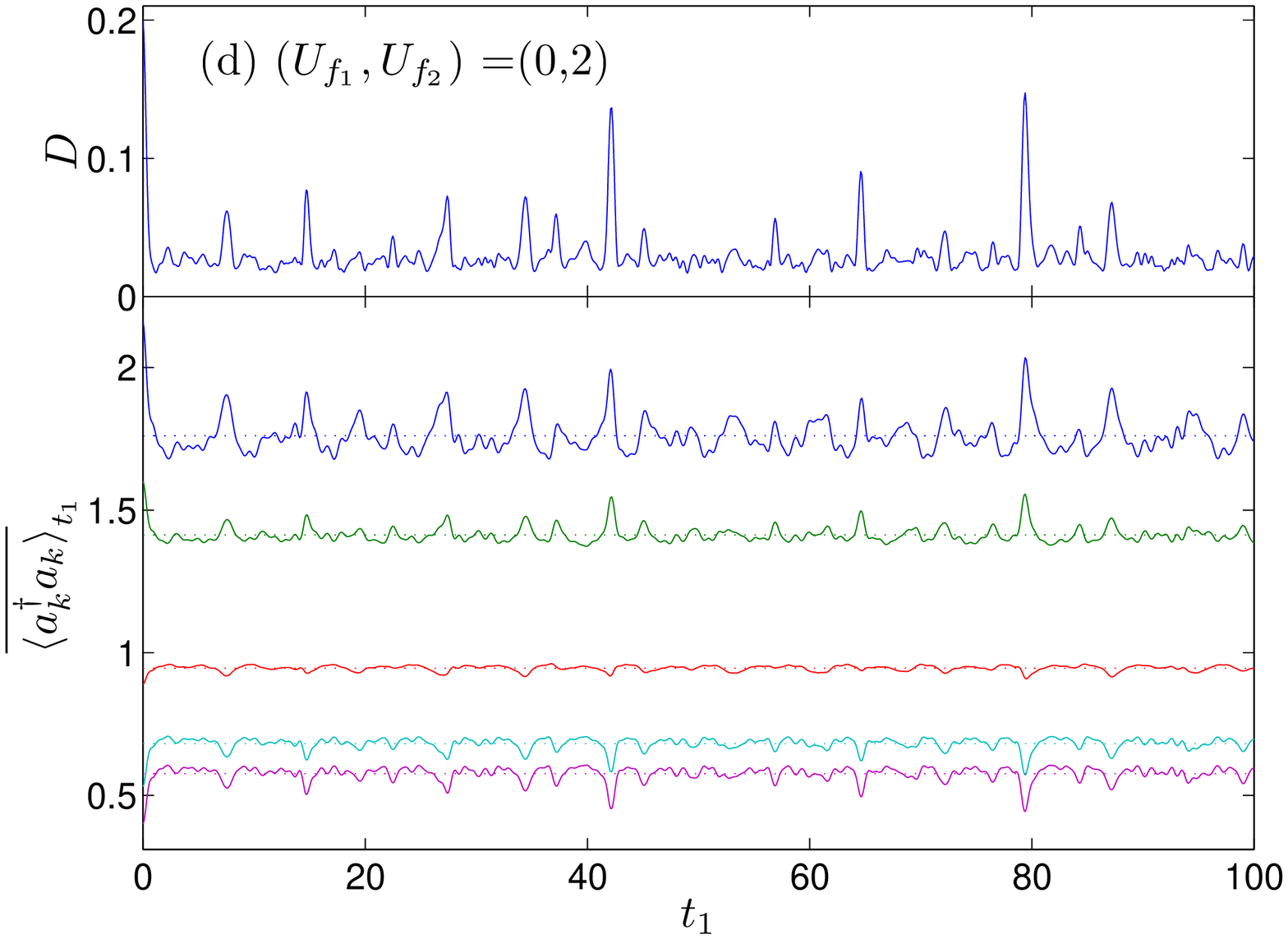}
\end{minipage}
\begin{minipage}[b]{0.24 \textwidth}
\centering
\includegraphics[ width=\textwidth]{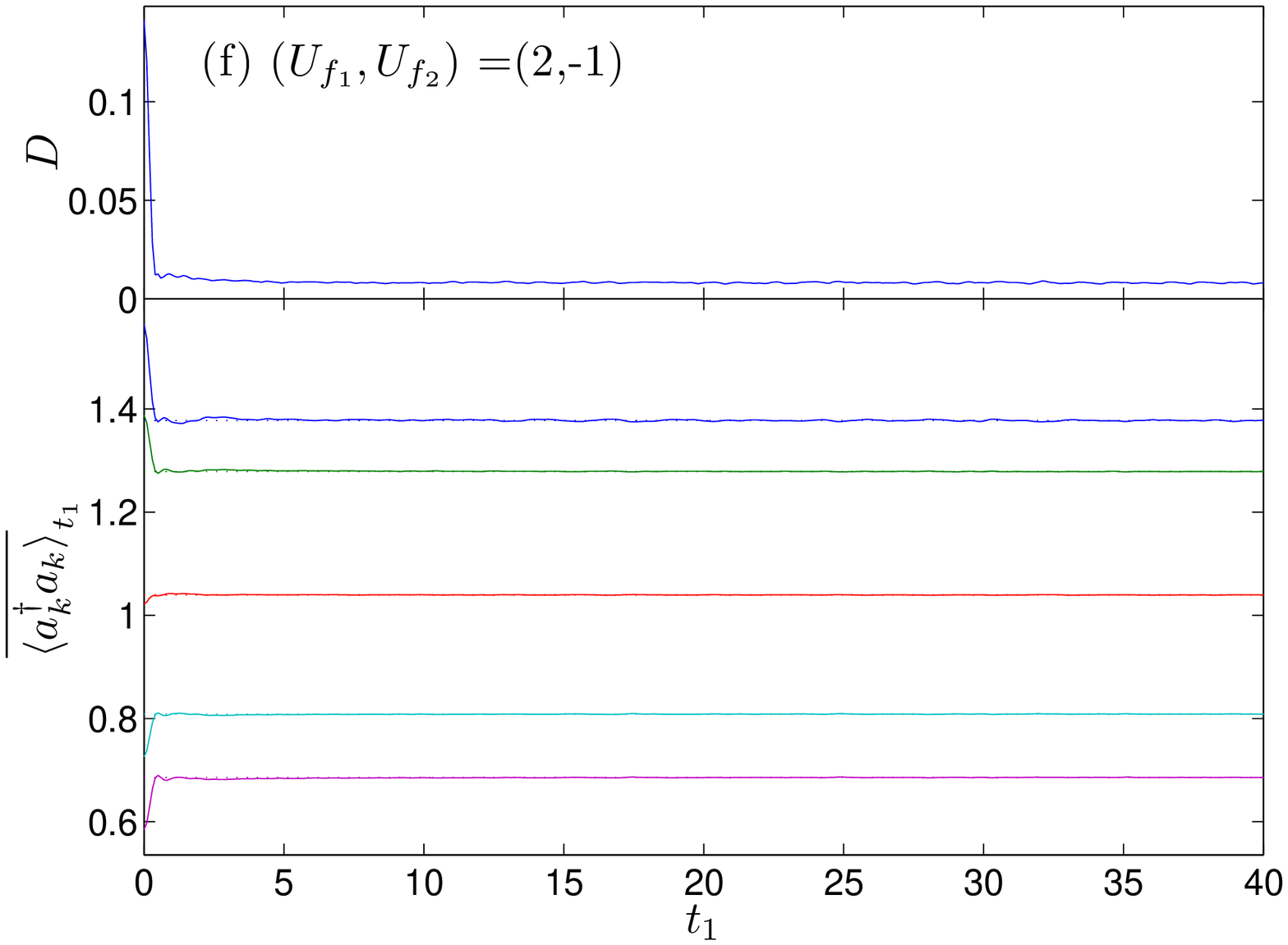}
\end{minipage}
\begin{minipage}[b]{0.24 \textwidth}
\centering
\includegraphics[ width=\textwidth]{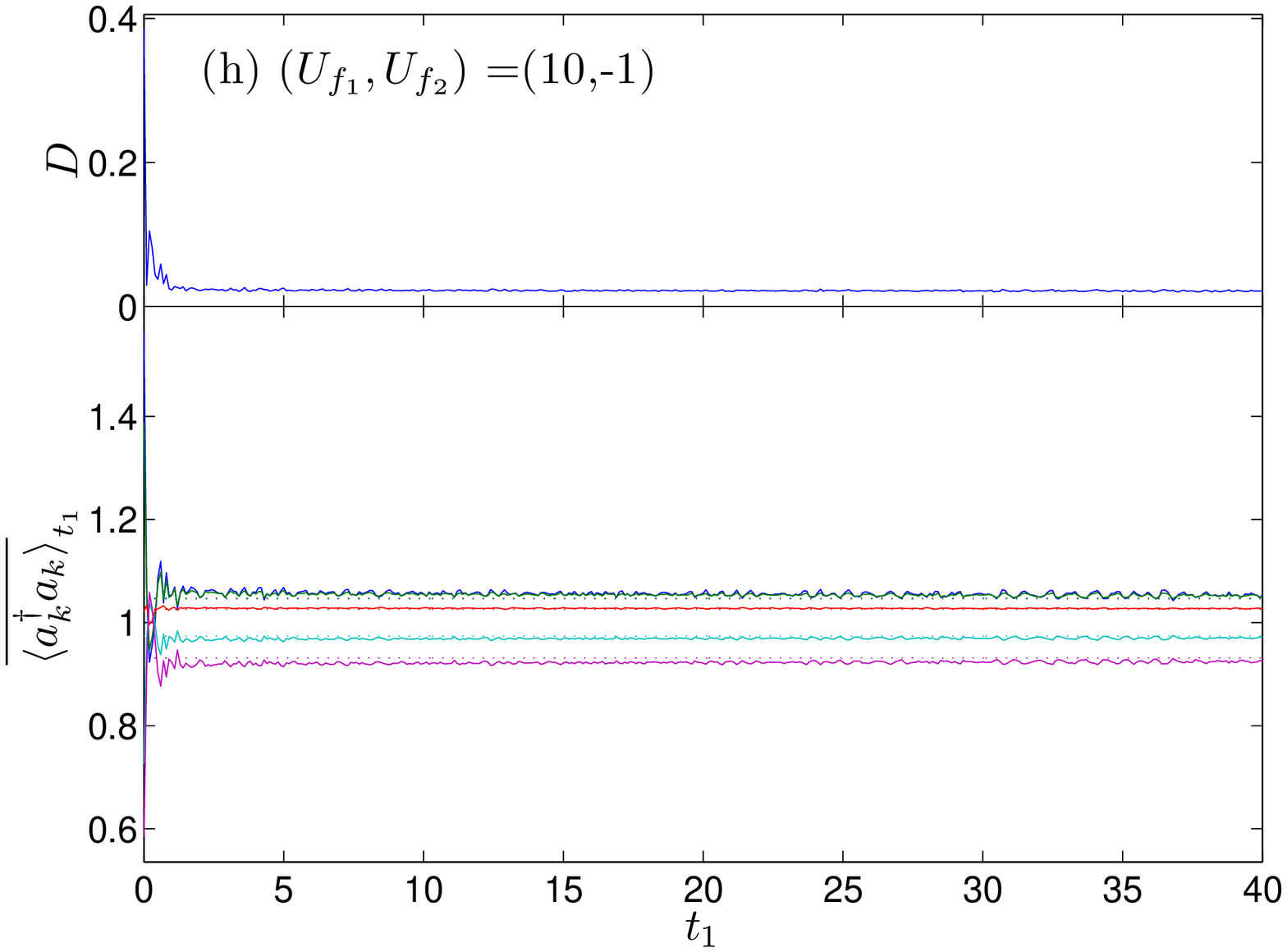}
\end{minipage}
\caption{(Color online) The distance $D $ between the
matrices $\Omega$ and $\bar{\rho}_{t_1}$ (upper panels) and
the time-averaged values of $\langle a^\dagger_k a_k
\rangle $ (lower panels), as functions of the time of the
second quench $t_1$. The dashed lines in the lower panels
indicate the average values of the corresponding solid
lines, i.e., values given by $\Omega$ [see
Eq.~(\ref{omega})]. The initial state is the same as in
previous Figures. The parameters $(U_{f_1},U_{f_2})$ are
shown in the inserts. \label{fig5}}
\end{figure*}

\begin{figure*}[tb]
\begin{minipage}[b]{0.24 \textwidth}
\centering
\includegraphics[ width=\textwidth]{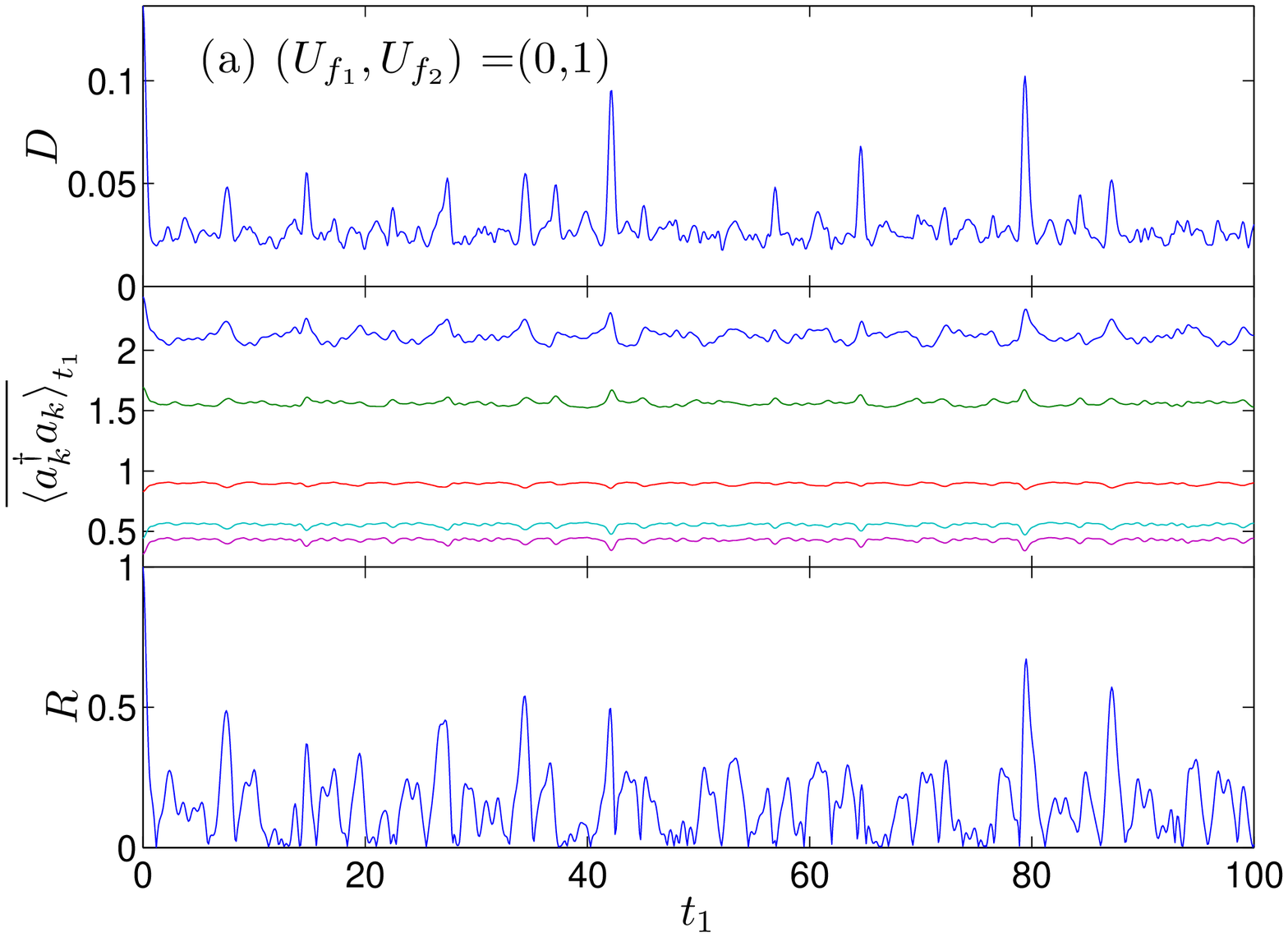}
\end{minipage}
\begin{minipage}[b]{0.24 \textwidth}
\centering
\includegraphics[ width=\textwidth]{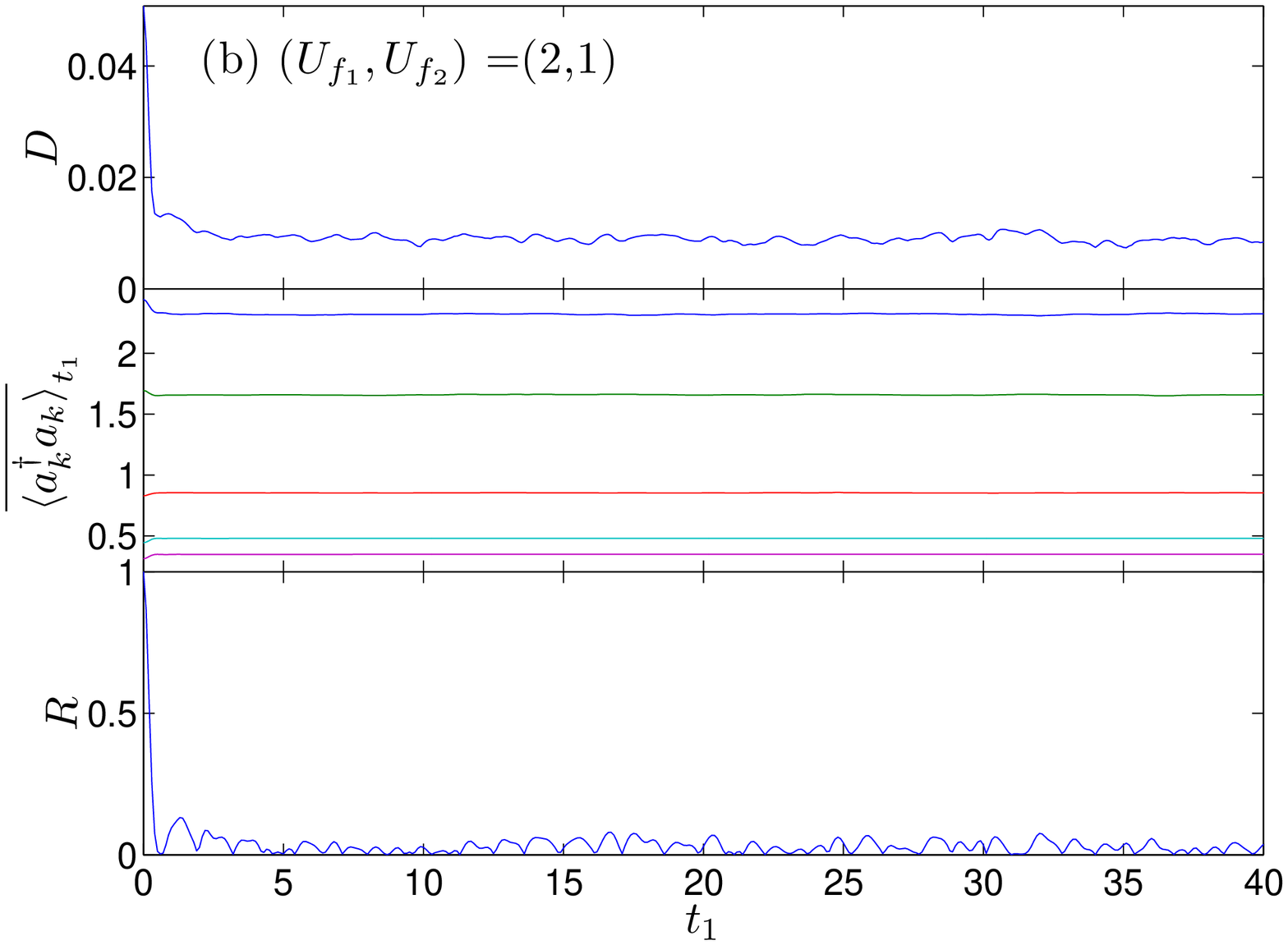}
\end{minipage}
\begin{minipage}[b]{0.24 \textwidth}
\centering
\includegraphics[ width=\textwidth]{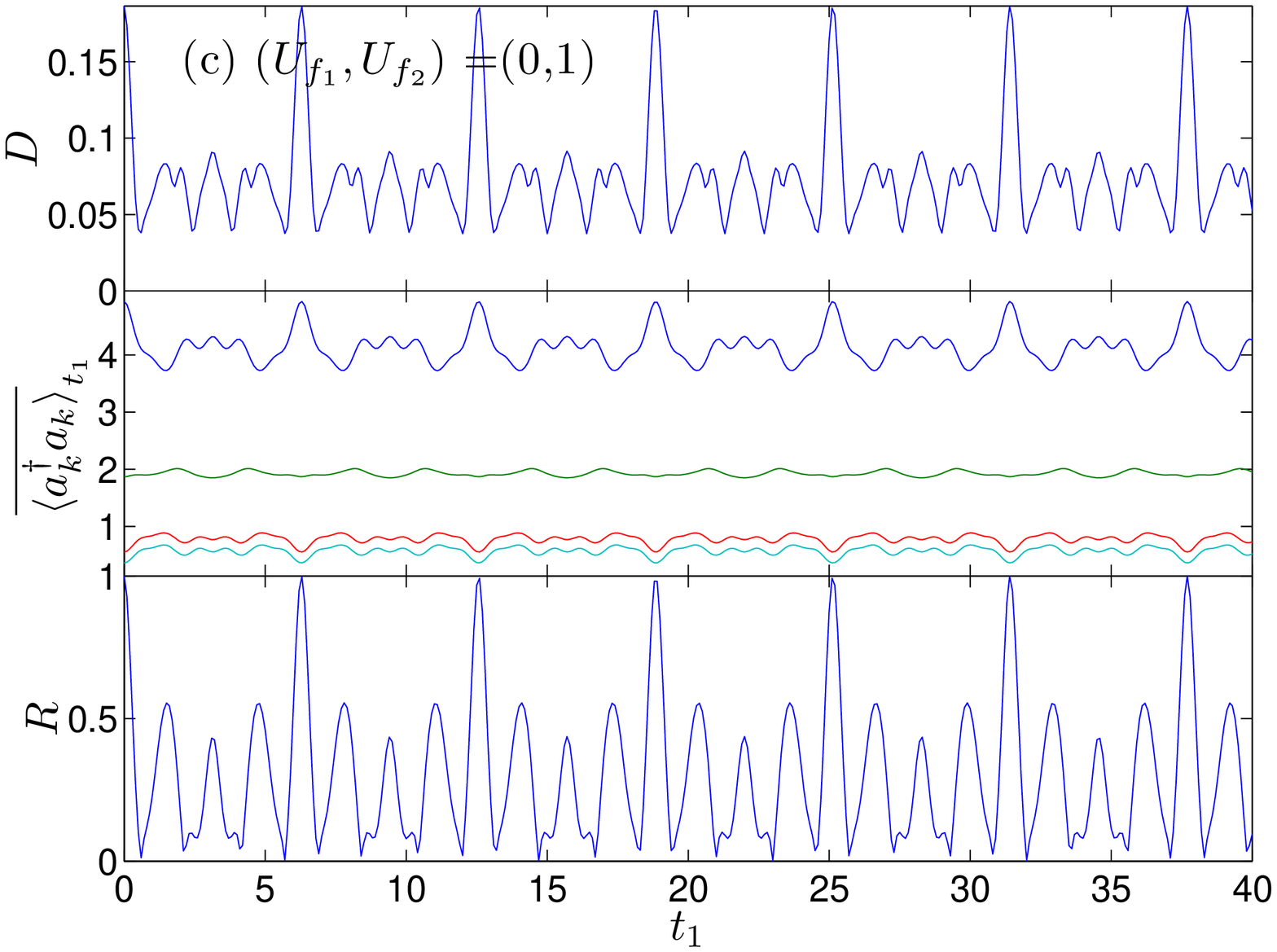}
\end{minipage}
\begin{minipage}[b]{0.24 \textwidth}
\centering
\includegraphics[ width=\textwidth]{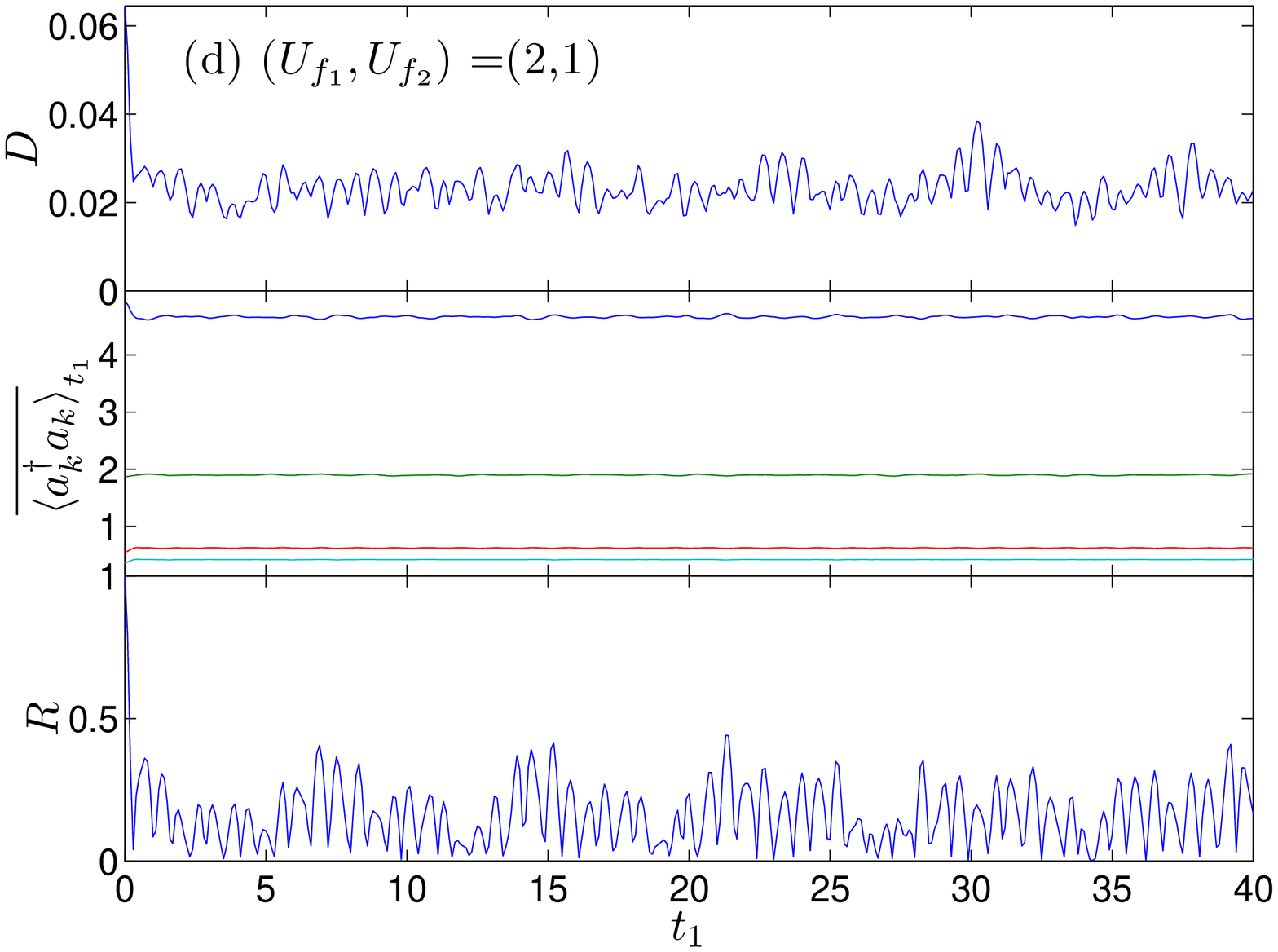}
\end{minipage}
\caption{(Color online) The figure-of-merit of recurrence
$R$ as a function of time. Also shown are
$D(\Omega,\bar{\rho}_{t_1})$ and $\overline{\langle
a^\dagger_k a_k \rangle }_{t_1} $. Note the correlation
between the three in (a) and (c). In (a) and (b), the
initial state is the same as in previous Figures, i.e.,
$(M,N,q,D_q)=(9,9,1,2700)$, $U_i=1$, and $\beta_i=0.3$. In
(c) and (d), the initial state is of
$(M,N,q,D_q)=(6,10,1,497)$, $U_i=1$, and $\beta_i=0.3$. The
values of $(U_{f_1},U_{f_2})$ are given in the inserts.
\label{fig6}}
\end{figure*}

\begin{figure*}[tb]
\begin{minipage}[b]{0.24 \textwidth}
\centering
\includegraphics[ width=\textwidth]{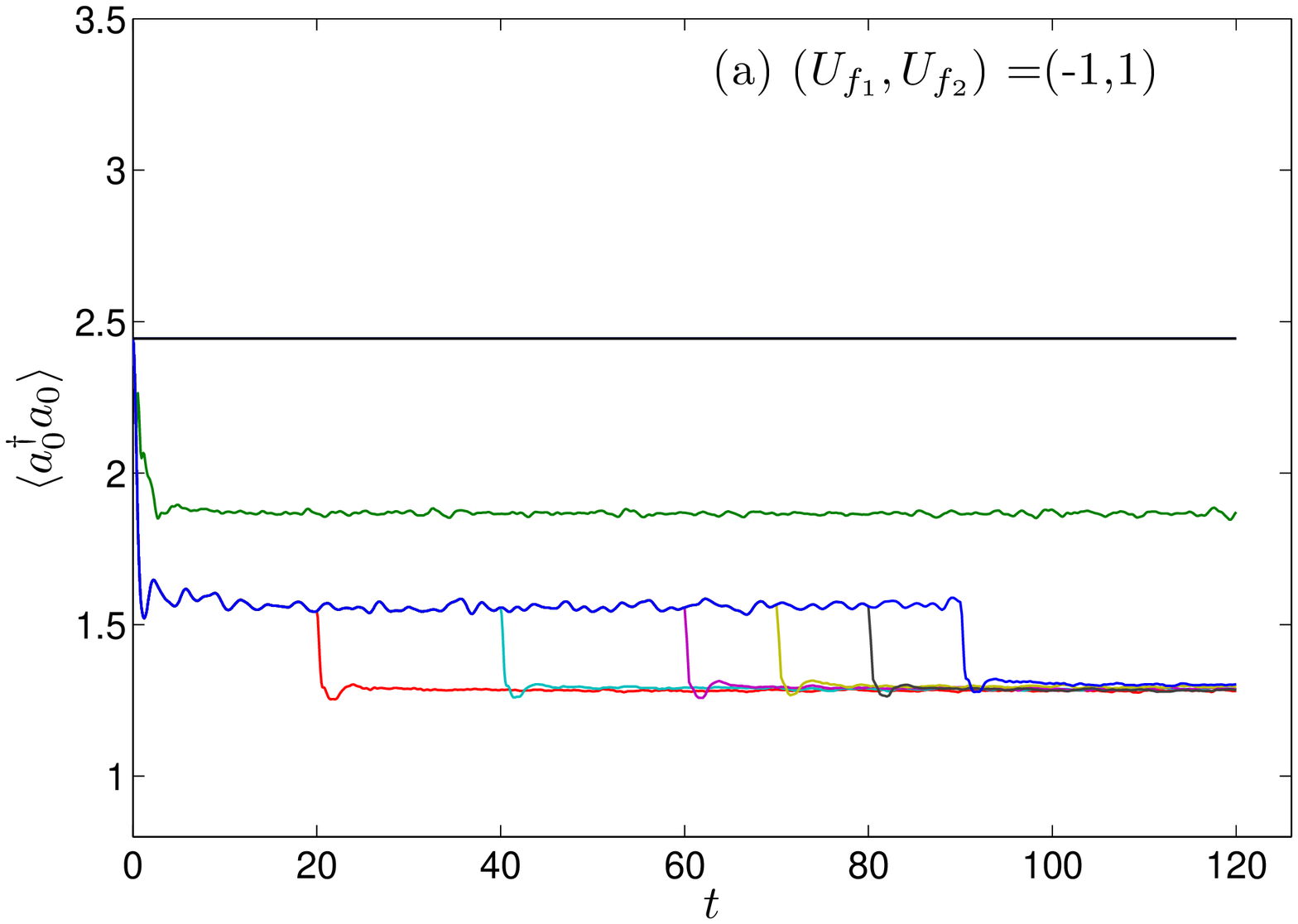}
\end{minipage}
\begin{minipage}[b]{0.24 \textwidth}
\centering
\includegraphics[ width=\textwidth]{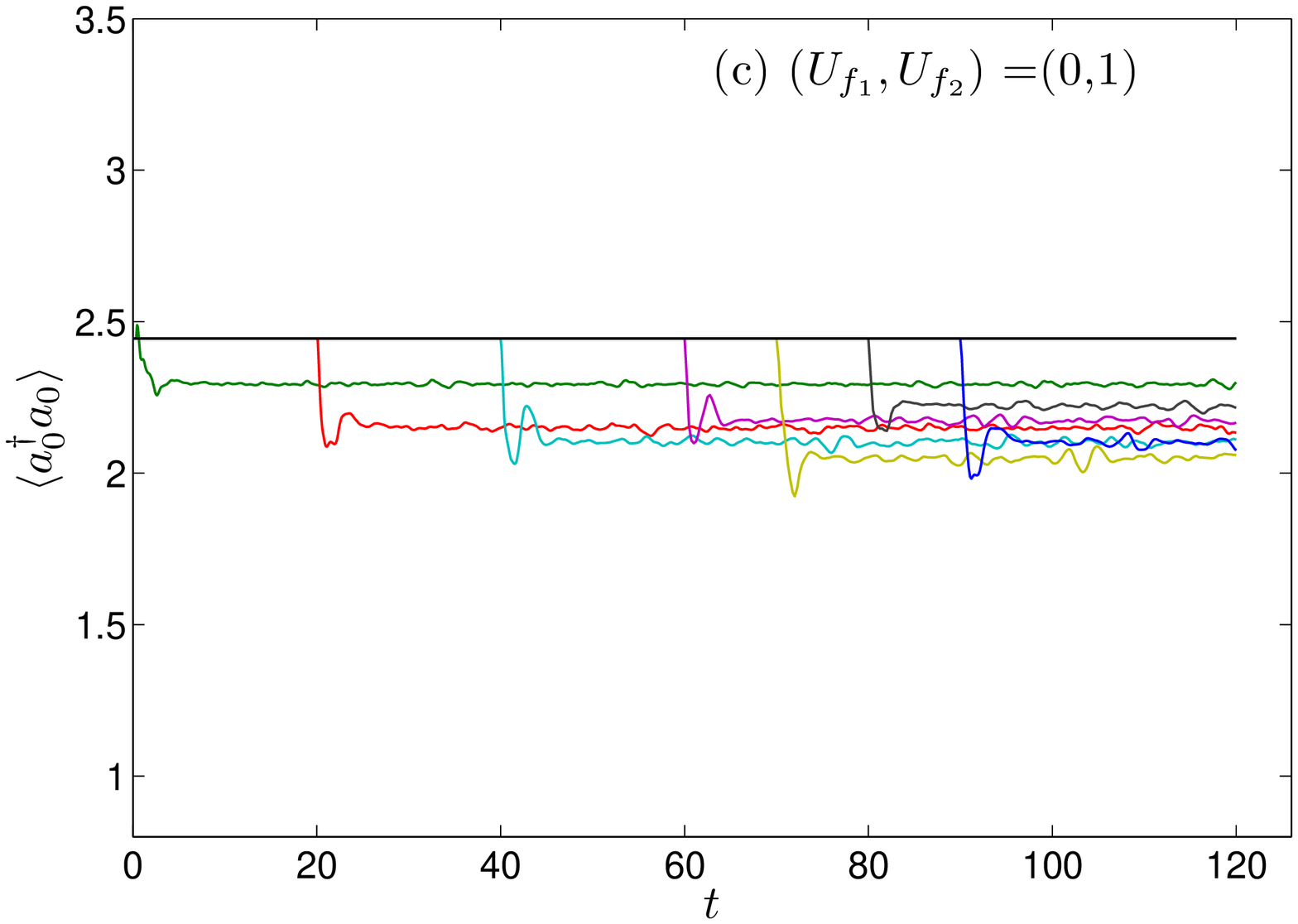}
\end{minipage}
\begin{minipage}[b]{0.24 \textwidth}
\centering
\includegraphics[ width=\textwidth]{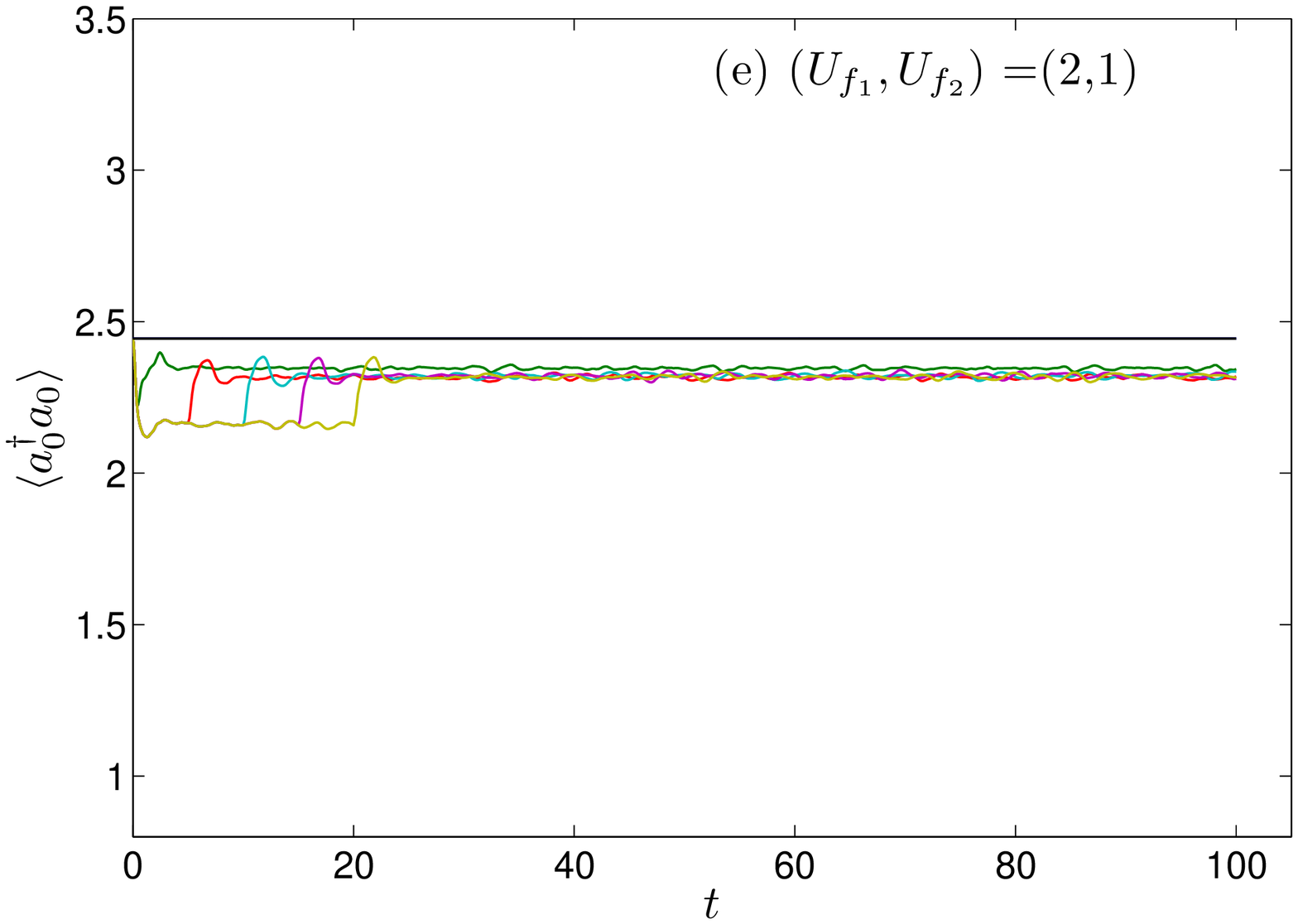}
\end{minipage}
\begin{minipage}[b]{0.24 \textwidth}
\centering
\includegraphics[ width=\textwidth]{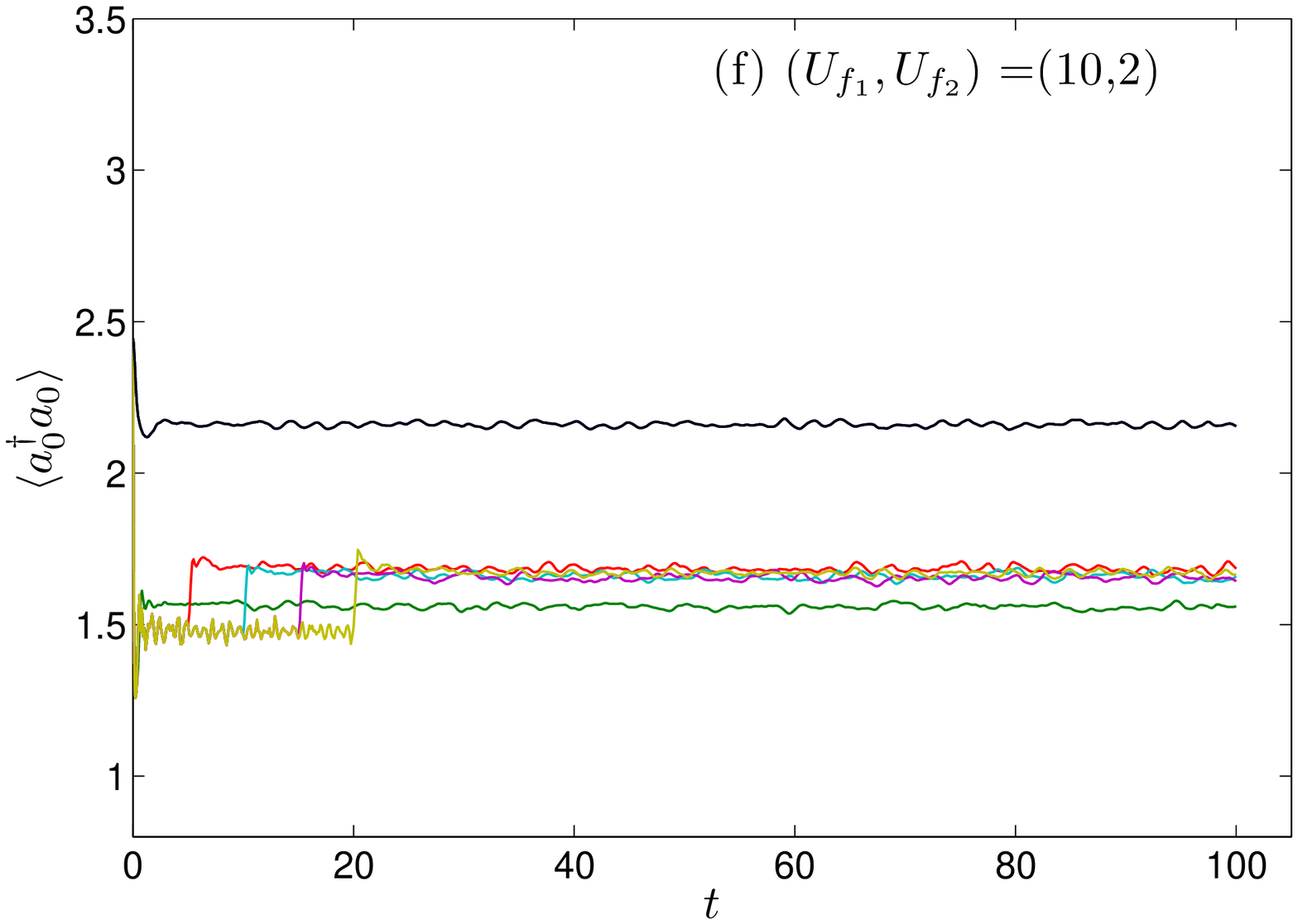}
\end{minipage}

\begin{minipage}[b]{0.24 \textwidth}
\centering
\includegraphics[ width=\textwidth]{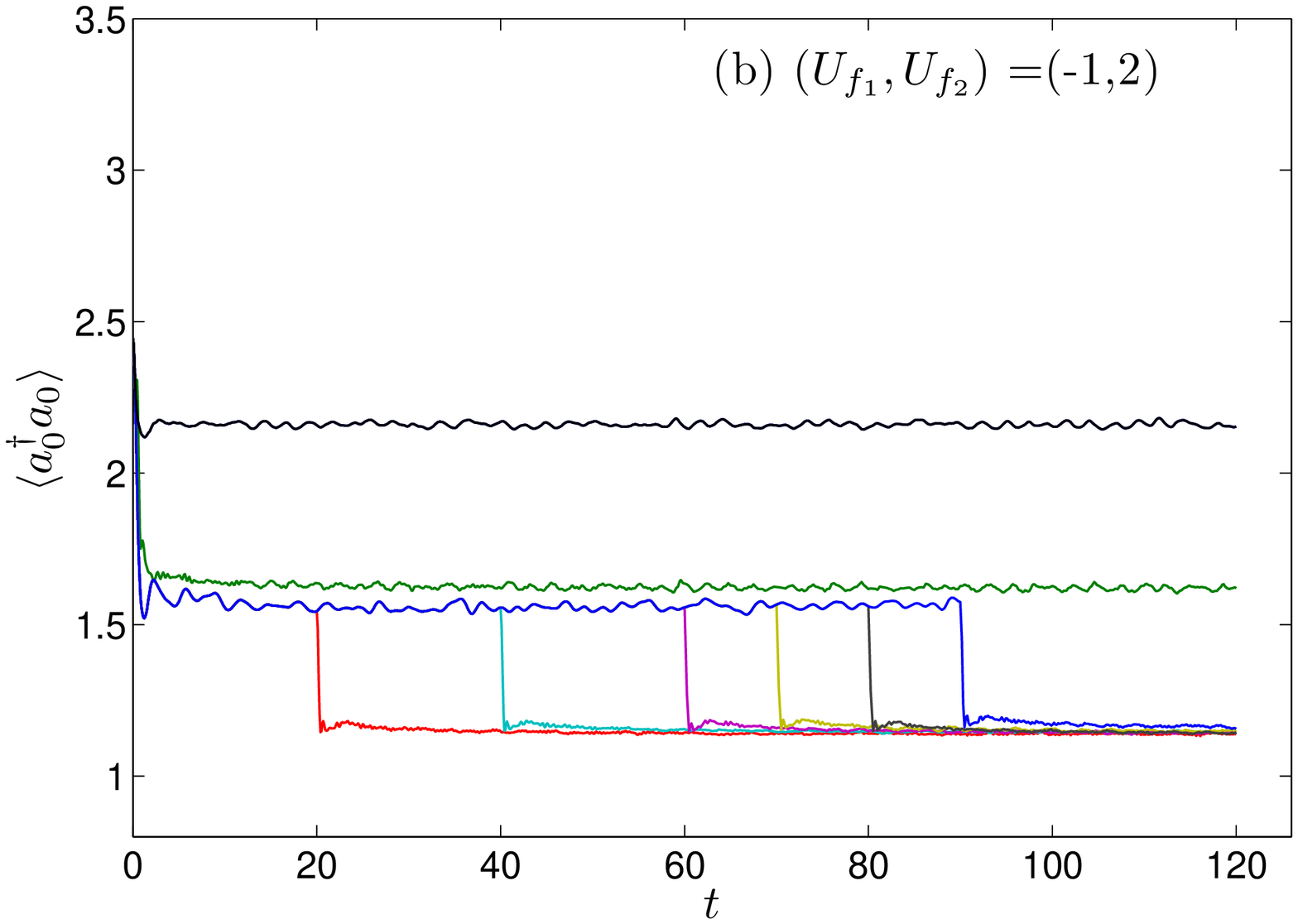}
\end{minipage}
\begin{minipage}[b]{0.24 \textwidth}
\centering
\includegraphics[ width=\textwidth]{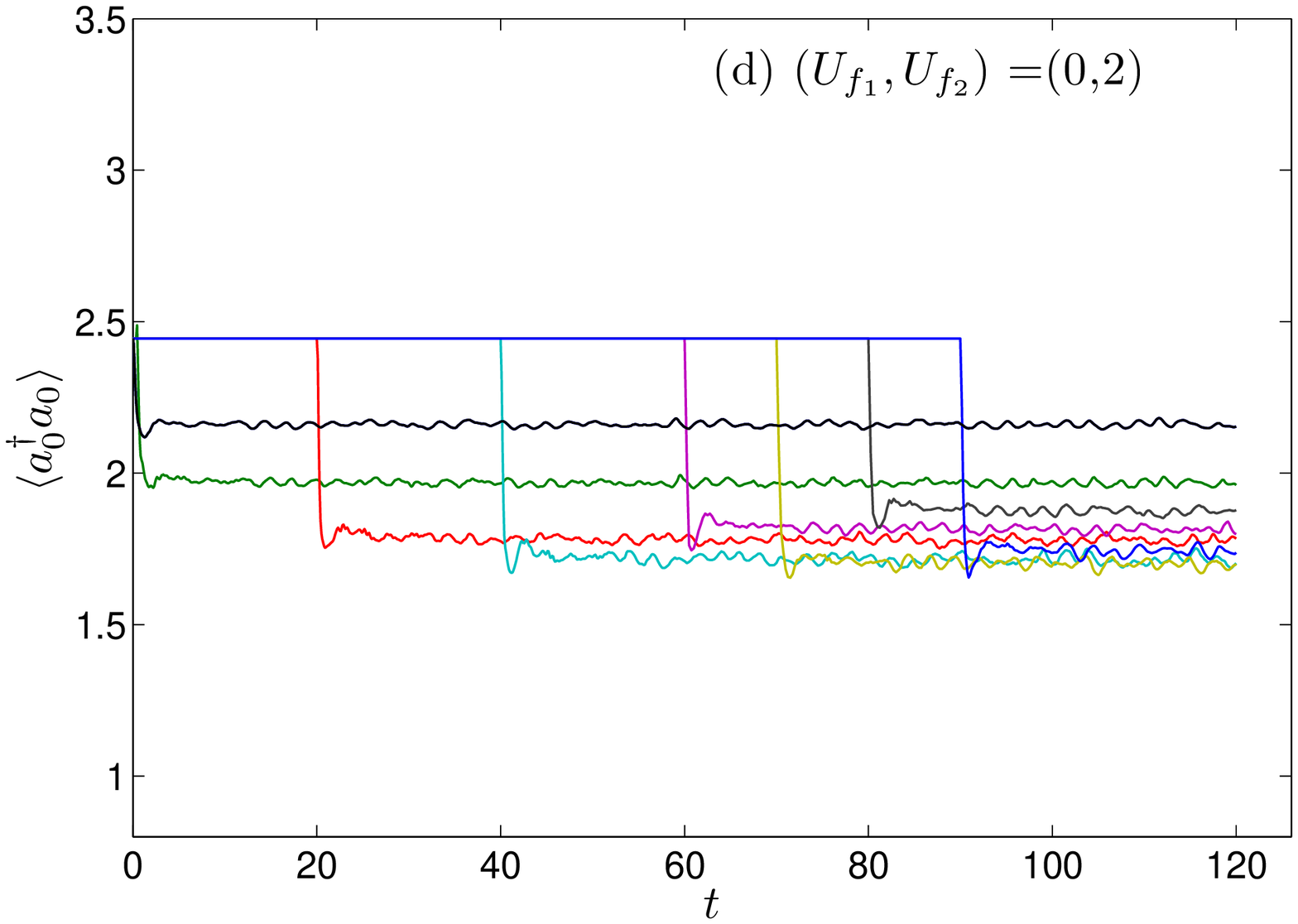}
\end{minipage}
\begin{minipage}[b]{0.24 \textwidth}
\centering
\includegraphics[ width=\textwidth]{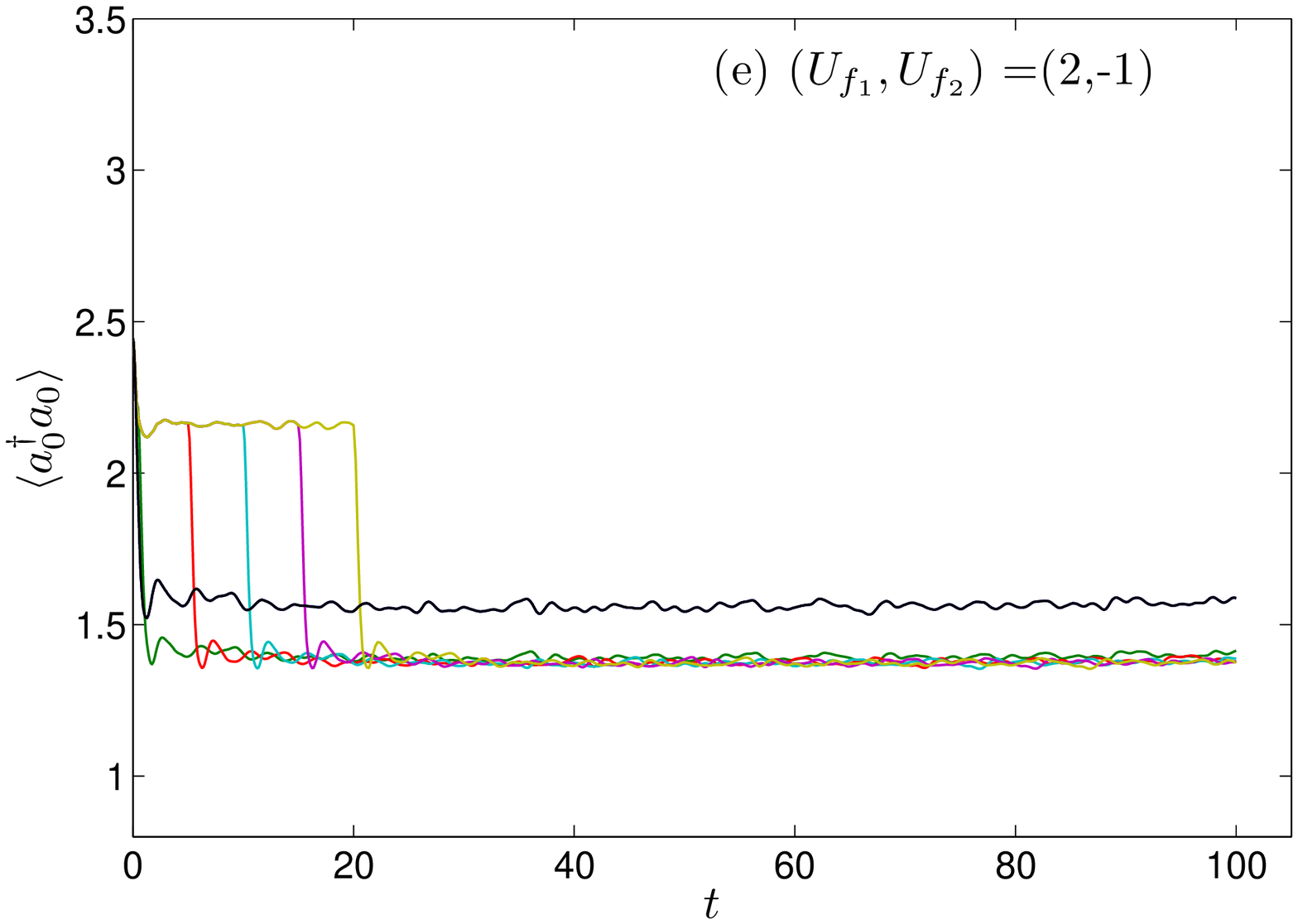}
\end{minipage}
\begin{minipage}[b]{0.24 \textwidth}
\centering
\includegraphics[ width=\textwidth]{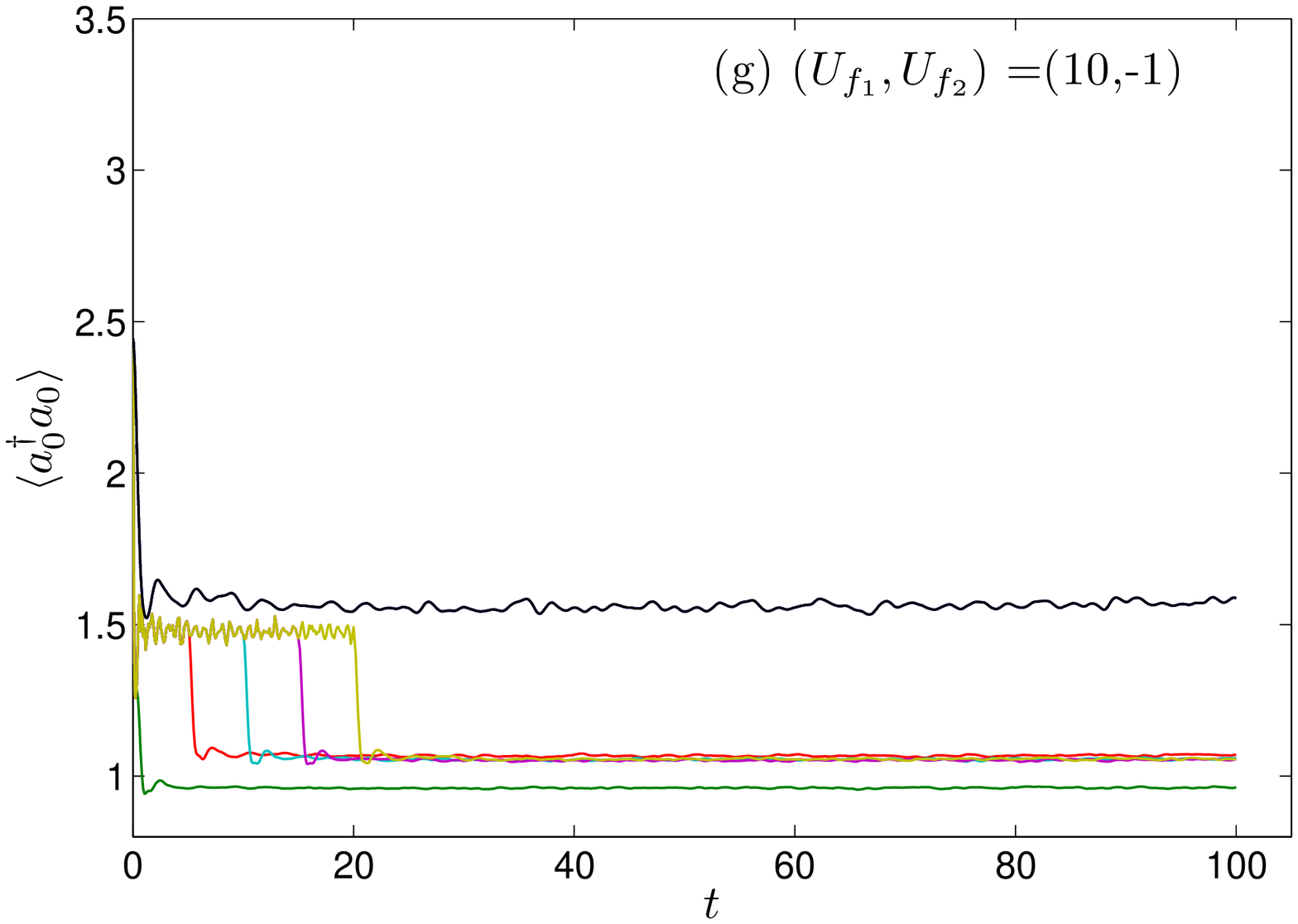}
\end{minipage}
\caption{(Color online) Time evolution of the population on
the $k=0$ Bloch state $\langle a_0^\dagger a_0 \rangle $
\cite{test}. Other $k$'s show similar behavior and thus are
not shown. The figures correspond to those in
Fig.~\ref{fig5} one-to-one. The initial state is the same
as in Fig.~\ref{fig1} and Fig.~\ref{fig4}. In (a)-(d), the
different $t_1$'s investigated are
$(0,0.35,20,40,60,70,80,90)$, while in (e)-(h), the
different $t_1$'s are $(0,0.35,5,10,15,20)$. Note that in
each figure, the black and green lines correspond to
$t_1=0$ and $0.35$, respectively. \label{fig7}}
\end{figure*}

It is shown in Fig.~\ref{fig4} that after a finite
transient time, the physical variables equilibrate to their
average values exhibiting minimal fluctuations. Moreover,
it has been proven that the amplitudes of the fluctuations
will decrease exponentially with the size of the system.
Therefore, the observation is that the system, described by
the density matrix $\rho(t)$, is almost indistinguishable
from a system described by the time-averaged density matrix
$\bar{\rho}$, as far as the simple realistic physical
variables are concerned. This is remarkable. Because though
$\rho(t)$ evolves unitarily and suffers no loss of
information of $\rho_i$, it behaves as if it were fully
decoherenced. The question is then, to what extent can we
hold onto this proposition? Is it possible to distinguish
$\rho(t)$ and $\bar{\rho}$, or $\rho(t_1)$ and $\rho(t_2)$
($t_1\neq t_2$), by some means? Motivated by this problem,
we have considered the scenario of giving the quenched
system a second quench. That is, after the first quench at
$t=0$ which changes $U$ from $U_i$ to $U_{f_1}$, at time
$t=t_1$, the system is quenched again by changing the value
of $U$ from $U_{f_1}$ to $U_{f_2}$, which is then held on
forever. The concern is, would the long-time dynamics of
the system depends on the specific time $t_1$?

Denote the Hamiltonian associated with $U_{f_2}$ as
$H_{f_2}$. The density matrix of the system later is given
by $\rho(t )=e^{-i H_{f_2} (t-t_1)} \rho(t_1) e^{i H_{f_2}
(t-t_1)} $ (for $t>t_1$). As before, we are interested in
the long-time averaged value of $\rho(t)$,
\begin{equation}
\bar{\rho}_{t_1}= \lim_{T\rightarrow \infty}  \frac{1}{T}\int_0^T dt \rho(t_1+t),
\end{equation}
since it has been shown and proven above that the dynamics
of the system is to a large extent captured by the
time-averaged density matrix. Here the subscript indicates
the dependence on the time $t_1$. It is also useful to
define the average of $\bar{\rho}_{t_1}$ with respect to
$t_1$,
\begin{eqnarray}
\Omega &=& \lim_{T\rightarrow \infty}\frac{1}{T} \int_0^T dt_1 \bar{\rho}_{t_1} \nonumber \\
&=& \lim_{T\rightarrow \infty}\frac{1}{T} \int_0^T dt  e^{-i H_{f_2} t}  \bar{\rho} e^{i H_{f_2} t}.
\end{eqnarray}
The second equality means that $\Omega$ is actually the
time-averaged density matrix associated with an initial
state $\bar{\rho}$ [see Eqs.~(\ref{def_bar_rho}) and
(\ref{diag})] and a Hamiltonian $H_{f_2}$. One purpose of
defining $\Omega$ is to set a reference state independent
of $t_1$.

To gain an overall idea of the dependence on $t_1$ of the
long-time dynamics, we have studied the distance between
$\bar{\rho}_{t_1}$ and $\Omega$ \cite{chuang}, and the
time-averaged value of $\langle a_k^\dagger a_k \rangle$,
\begin{eqnarray}
\overline{\langle a^\dagger_k a_k \rangle }_{t_1} &\equiv & \lim_{T\rightarrow \infty}\frac{1}{T} \int_0^T dt  \cdot tr(\rho(t_1+t) a^\dagger_k a_k) \nonumber \\
&=& tr(\bar{\rho}_{t_1} a^\dagger_k a_k),
\end{eqnarray}
as functions of $t_1$. Note that the average value of
$\overline{\langle a^\dagger_k a_k \rangle }_{t_1}$ with
respect to $t_1$ is given by $\Omega$,
\begin{equation} \label{omega}
\lim_{T\rightarrow \infty} \frac{1}{T}\int_0^T d t_1
\overline{\langle a^\dagger_k a_k \rangle }_{t_1}=
tr(\Omega  a^\dagger_k a_k ).
\end{equation}
This is another reason for defining $\Omega$. The
quantities $D(\Omega,\bar{\rho}_{t_1})$ and
$\overline{\langle a^\dagger_k a_k \rangle }_{t_1} $ are
shown in Fig.~\ref{fig5}. Eight pairs of
$(U_{f_1},U_{f_2})$ are examined with the same initial
condition as in Fig.~\ref{fig1}. We see that for all cases
with $U_{f_1}\neq 0$, both $D$ and $\overline{\langle
a^\dagger_k a_k \rangle }_{t_1} $ set down to their average
values quickly. However, for the special case of
$U_{f_1}=0$, both $D$ and $\langle a_k^\dagger a_k
\rangle_{t_1}$ display repeated recurrences, without any
sign of equilibration. The situation is the reverse of that
in Fig.~\ref{fig4}, where $\langle a_k^\dagger a_k \rangle
$ does not show any fluctuations in the case of
$U_{f_1}=0$.

This phenomenon is due to the recurrence of the density
matrix $\rho(t)$ to $\rho_i$ \cite{recur}. From
Eq.~(\ref{rhot}), we see that in the representation of
$\{|\psi^{f_1}_m\rangle \}$, the $mn$-th off-diagonal
element of $\rho(t)$ rotates at an angular frequency of
$E^{f_1}_m-E^{f_1}_n$. In the generic case of $U_{f_1}\neq
0$, the energy gaps $E^{f_1}_m-E^{f_1}_n$ are quite random
and incommensurate, and thus recurrence of the density
matrix is rare. More precisely, the span between two times
when all the matrix elements of $\rho(t)$ get (nearly) in
phase again is extraordinarily large. On the contrary, in
the special case of $U_{f_1}=0$, all eigenvalues and hence
all the energy gaps $E^{f_1}_m-E^{f_1}_n$ are integral
combinations of the few basic frequencies $\omega_k$, and
thus the probability of recurrence is much higher. To
demonstrate that the sharp peaks in Figs.~\ref{fig5}c and
\ref{fig5}d are due to recurrences of the density matrix
$\rho(t)$ to $\rho_i$, we define the figure-of-merit of
recurrence,
\begin{equation}
R (t)= \frac{|\sum_{m,n}' \rho_{m,n}^2 e^{-i(E^{f_1}_m-E^{f_1}_n) t}|}{\sum_{m,n}' \rho_{mn}^2},
\end{equation}
where the prime means the summation is over $(m,n)$ such
that $E^{f_1}_m\neq E^{f_1}_n $. It is clear that $0\leq R
\leq 1$ and $R=1$ when and only when all the off-diagonal
elements get in phase.

In Figs.~\ref{fig6}a and \ref{fig6}b, which share the same
parameters as Figs.~\ref{fig5}c and \ref{fig5}e
respectively, we have shown $R(t_1)$ together with
$D(\Omega,\bar{\rho}_{t_1})$ and $\overline{\langle
a^\dagger_k a_k \rangle }_{t_1} $. In Fig.~\ref{fig6}a, we
see that every time $D(\Omega,\bar{\rho}_{t_1})$ and
$\overline{\langle a^\dagger_k a_k \rangle }_{t_1} $ get
close to their values at $t_1=0$, $R(t_1)$ shows a peak. In
other words, there is a strong positive correlation between
$R(t_1)$ and $D(\Omega,\bar{\rho}_{t_1})$ and
$\overline{\langle a^\dagger_k a_k \rangle }_{t_1} $. In
comparison, in Fig.~\ref{fig6}b, $R(t_1)$ drops quickly
from unity to less than $0.2$ and remains low all the time,
and in turn $D(\Omega,\bar{\rho}_{t_1})$ and
$\overline{\langle a^\dagger_k a_k \rangle }_{t_1}  $ do
not show any recurrence. To further consolidate the
connection between the recurrence of $\rho(t)$ and that of
$D(\Omega,\bar{\rho}_{t_1})$ and $\overline{\langle
a^\dagger_k a_k \rangle }_{t_1} $, we have considered the
case of $M=6$. In this case, if $U_{f_1}=0$, all the basic
frequencies $\omega_k$ are commensurate, and thus there
exist perfect recurrences, as shown in Fig.~\ref{fig6}c.
There we see clearly that $D(\Omega,\bar{\rho}_{t_1})$ and
$\overline{\langle a^\dagger_k a_k \rangle }_{t_1} $ return
to their original values at $t_1=0$ periodically, and this
happens when and only when $R$ returns to unity. However,
once $U_{f_1} \neq 0$ is set nonzero (see Fig.~\ref{fig6}d)
and thus the commensurability of the energy gaps is
destroyed, the situation returns to that in
Fig.~\ref{fig6}b.

The fact revealed in Fig.~\ref{fig5} and Fig.~\ref{fig6} is
quite interesting. The long-time dynamics of the system  is
sensitive or insensitive to the exact time when the second
quench is applied, depending on whether the intermediate
Hamiltonian $H_{f_1}$ is integrable ($U_{f_1}=0$) or
non-integrable ($U_{f_1}\neq 0$). In the integrable case,
$\overline{\langle a^\dagger_k a_k \rangle }_{t_1} $
exhibits large fluctuations and repeated recurrences. The
system retains the memory of the initial state under the
control of the Hamiltonian $H_{f_1}$. By contrast, in the
non-integrable case, $\overline{\langle a^\dagger_k a_k
\rangle }_{t_1} $ go over to their average values
(predicted by $\Omega$) after a transitory period, showing
little dependence on $t_1$ afterwards. Combined with
Fig.~\ref{fig4}, the picture is that $\rho(t)$ evolving
under the control of a non-integrable Hamiltonian, not only
yields the expectation values of $a_k^\dagger a_k$ as if it
were $\bar{\rho}$, but even responds to the second quench
as if it were $\bar{\rho}$.

In Fig.~\ref{fig7}, we have checked this picture by
studying the real time evolution of $\langle a_k^\dagger
a_k \rangle$ with $k=0$ under the double-quench scenario.
The eight figures shown correspond to those in
Fig.~\ref{fig5} respectively. For each pair of
$(U_{f_1},U_{f_2})$, we have studied the evolution of
$\langle a_0^\dagger a_0 \rangle$ for several different
values of $t_1$. We see that in all the cases with
$U_{f_1}\neq 0$, as long as $t_1$ is larger than the
transient time, which can be roughly read from
Fig.~\ref{fig5}, the later evolution of $\langle
a_0^\dagger a_0 \rangle$ is quantitatively independent of
$t_1$. On the contrary, in the case with $U_{f_1}=0$, the
later values of $\langle a_0^\dagger a_0 \rangle $ vary
wildly for different values of $t_1$.

Here it is instructive to combine Fig.~\ref{fig4} and
Fig.~\ref{fig7} and compare. In the $U_{f_1}\neq 0$ cases,
there is a sense of typicality \cite{goldstein,popescu}.
The density matrix $\rho(t)$ governed by $H_{f_1}$ is
surely non-stationary. However, for $\rho(t)$ at different
times, they yield almost the same expectation values for
the observables, and moreover, they share almost the same
response to the same quench. In the case of $U_{f_1}=0$,
what Fig.~\ref{fig7} reveals is a good complement to that
in Fig.~\ref{fig4}. It demonstrates that it is
inappropriate to say that the system thermalizes in this
case, even though the density matrices and expectation
values of the observables agree---since according to one's
everyday experience, a system in thermal equilibrium should
not show any time dependence.

\section{conclusions and discussions}\label{sec5}
We have studied the quench dynamics of the Bose-Hubbard
model both analytically and numerically. The issues of
\therm\ and equilibration are investigated thoroughly.

On the \therm\ side, which concerns whether the quenched
system behaves like a canonical ensemble, it is found that
this is the case only for small-amplitude quenches (at
least for the finite-sized system investigated). However,
the time-averaged density matrix does manifest many
interesting features in different regimes. These features
are self-consistently understood after a study of the
overlaps between the eigenstates of $H_i$ and $H_{f_1}$.
Here we would like to say that it is urgent and would be
very helpful to develop some analytical tools so that some
general relations between the eigen-systems of $H_i$ and
$H_{f_1}$ can be established. These tools and relations
would also be useful to determine whether the non-\therm\
phenomenon observed is just a finite-size effect.

On the equilibration side, which is about whether physical
observables relax to stationary values without appreciable
fluctuations, the result is that this is indeed the case
for quantities as $\langle a^\dagger_k a_k \rangle$ which
are of most interest. Moreover, it is proven analytically
that for these quantities the fluctuations in time will
decay exponentially with the size of the system. Therefore,
the overall picture is that generally the system
equilibrates but without \therm.

The second quench reveals something more subtle. First, the
subsequent dynamics depends or not on the second quench
time $t_1$ according to $U_{f_1}=0$ or not. The underline
reason is the recurrence or not of the initial density
matrix, which in turn has its root in the eigenvalue
statistics of the Hamiltonian $H_{f_1}$. This effect leaves
us the impression that a non-integrable Hamiltonian has
more ``dephasing power'' than an integrable one. Possibly
it can be a tool to check the integrability of a
Hamiltonian. Second, in the case of $U_{f_1}\neq 0$, it is
found that the system described by $\rho(t_1)$ responds to
the second quench as if it were $\bar{\rho}$ for $t_1$
larger than the transient time. This means that we can take
the equilibration more serious---$\rho(t_1)$ and
$\bar{\rho}$ not only yield almost the same expectation
values for the generic physical variables but also yield
almost the same dynamics after a quench. Moreover, the fact
that the transient time is short indicates that the
intermediate Hamiltonian $H_{f_1}$, which is
non-integrable, is effective in ``dephasing'' the initial
density matrix. In another perspective, the dynamics of the
system is sensitive to the fluctuations of $U$. This has
the implication that in future experiments, accurate
control of $U$ would be a necessity to interpret the
results correctly.

\section{acknowledgment}
We are grateful to H.~T. Yang, Z.~X. Gong, Y.-H. Chan,
L.~M. Duan, and D.~L. Zhou for stimulating discussions and
valuable suggestions. J. M. Z. is supported by NSF of China
under Grant No.~11091240226.


\begin{thebibliography}{99}
\bibitem{negative}
We note that the Bose-Hubbard model with a negative $U$ has
not a well-defined thermodynamic limit. However, this
should not prevent us from studying the negative-$U$ case
at finite sizes.
\bibitem{prod}
It is so called because $P_m(E)>0$ and $\int d E P_m(E)=1$.
\bibitem{linear}
This is surely not the case for $U_i<0$. But it is believed
that it is indeed the case for $U_i \geq 0$.
\bibitem{test}
Only the quantity $\langle a_0^\dagger a_0 \rangle $ is
shown because it constitutes the most strigent test.
\end{thebibliography}
\end{document}